\documentclass[12pt, twocolumn]{aastex63}

\newcommand{\kmsmpc}{\rm km~s^{-1}~Mpc^{-1}}
\newcommand{\sigbcg}{\sigma_{\rm *,~BCG}}
\newcommand{\sigbcgc}{\sigma_{\rm *,BCG,3~kpc}}
\newcommand{\sigbcgh}{\sigma_{\rm *,~BCG,~R_{half}}}
\newcommand{\sigcl}{\sigma_{\rm cl}}
\newcommand{\Msun}{{\rm M}_{\odot}}

\usepackage{color}
\usepackage{multirow}

\begin{document}

\title{IllustrisTNG Snapshots for 10 Gyr of Dynamical Evolution of Brightest Cluster Galaxies and Their Host Clusters}

\author{Jubee Sohn}
\affiliation{Smithsonian Astrophysical Observatory, 60 Garden Street, Cambridge, MA 02138, USA}
\affil{Astronomy Program, Department of Physics and Astronomy, Seoul National University, Gwanak-gu, Seoul 151-742, Republic of Korea}

\author{Margaret J. Geller}
\affiliation{Smithsonian Astrophysical Observatory, 60 Garden Street, Cambridge, MA 02138, USA}

\author{Mark Vogelsberger}
\affiliation{Department of Physics, Kavli Institute for Astrophysics and Space Research, Massachusetts Institute of Technology, Cambridge, MA 02139, USA}

\author{Josh Borrow}
\affiliation{Department of Physics, Kavli Institute for Astrophysics and Space Research, Massachusetts Institute of Technology, Cambridge, MA 02139, USA}

\email{jbsohn@astro.snu.ac.kr}

%========================================
\begin{abstract}
We explore the redshift evolution of the dynamical properties of massive clusters and their brightest cluster galaxies (BCGs) at $z < 2$ based on the IllustrisTNG-300 simulation. We select 270 massive clusters with $M_{200} < 10^{14}~\Msun$ at $z = 0$ and trace their progenitors based on merger trees. From 67 redshift snapshots covering $z < 2$, we compute the 3D subhalo velocity dispersion as a cluster velocity dispersion ($\sigcl$). We also calculate the 3D stellar velocity dispersion of the BCGs ($\sigbcg$). Both $\sigcl$ and $\sigbcg$ increase as universe ages. The BCG velocity dispersion grows more slowly than the cluster velocity dispersion. Furthermore, the redshift evolution of the BCG velocity dispersion shows dramatic changes at some redshifts resulting from dynamical interaction with neighboring galaxies (major mergers). We show that $\sigbcg$ is comparable with $\sigcl$ at $z > 1$, offering an interesting observational test. The simulated redshift evolution of $\sigcl$ and $\sigbcg$ generally agrees with an observed cluster sample for $z < 0.3$, but with large scatter. Future large spectroscopic surveys reaching to high redshift will test the implications of the simulations for the mass evolution of both clusters and their BCGs.
\end{abstract}
%========================================

%========================================
\section{INTRODUCTION}
%========================================

Brightest cluster galaxies (BCGs) are the most luminous galaxies in the universe usually found near the center of galaxy clusters. BCGs have several additional distinctive features including large sizes and large velocity dispersions even compared to similarly massive galaxies that are not resident in rich clusters \citep{vonderLinden07, Bernardi09}. The BCG formation process could thus have features that are distinctive from other galaxies. 

The special location of BCGs suggests that BCG evolution is tightly connected with cluster evolution. In the hierarchical structure formation model, galaxy clusters grow through stochastic accretion of less massive systems (e.g., \citealp{vandenBosch02, McBride09, Zhao09, Fakhouri10, Kravtsov12, DeBoni16, Pizzardo21, Pizzardo22}). BCGs grow along with their host clusters. BCGs grow through active accretion of other, generally less massive, cluster members and intracluster material. Baryonic physics also plays an important role in BCG evolution. Thus, the study of the co-evolution of clusters and BCGs offers tests of both structure and galaxy formation models.

Many observational studies investigate the co-evolution of clusters and their BCGs (e.g., \citealp{Lin04, OlivaAltamirano14, Lin17, Loubser18, Kravtsov18, Wen18, Erfanianfar19, DeMaio20, Jung22}). In particular, a large number of cluster samples provide the stellar mass to halo mass relation (e.g., \citealp{Moster10, Kravtsov18, Erfanianfar19}). The observed mass ratio between clusters and their BCGs decreases as a function of cluster mass for cluster masses $> 10^{13}~\Msun$ (e.g., \citep{Erfanianfar19, Girelli20}). The stellar mass to halo mass relation suggests that BCG mass growth is suppressed in more massive cluster halos presumably by strong feedback processes including active galactic nuclei (AGN). This observed stellar mass to halo mass relation agrees well with the theoretical prediction (e.g., \citealp{RagoneFigueroa18, Behroozi19}). 

\citet{Sohn20} and \citet{Sohn21} explore the co-evolution of clusters and their BCGs based on the cluster velocity dispersion ($\sigcl$) and the stellar velocity dispersion of the BCGs ($\sigbcg$) (see also \citealp{Lauer14, Kim17, Loubser18}). The dynamical scaling relations provide independent tools for studying the co-evolution of clusters and BCGs. Furthermore, the velocity dispersions are insensitive to systematic biases that may affect the photometric measurements (i.e., luminosity and stellar mass). The observed $\sigbcg / \sigcl$ ratio decreases as a function of $\sigcl$, consistent with the stellar mass to halo mass relation. 

The observed velocity dispersion scaling relation is generally consistent with the theoretical predictions derived from numerical simulations. \citet{Marini21} showed that the $\sigbcg \- \sigcl$ relation derived from the DIANOGA simulations agrees with the observed relation (see also \citealp{Dolag10, Remus17}). \citet{Sohn22} use the IllustrisTNG-300 simulation that includes 280 massive cluster halos ($M_{200} > 10^{14}~\Msun$) to derive the $\sigbcg \- \sigcl$ relation. Again the relations from observations and simulations are generally consistent even when more simulated clusters with low $\sigbcg$ are included. 

Cosmological numerical simulations enable identification of clusters and BCGs with their progenitors (e.g., \citealp{Vogelsberger14a, Vogelsberger20, Nelson19}). \citet{RagoneFigueroa18} investigate the stellar mass assembly of BCGs at $z < 4$. \citet{Marini21} explore the evolution of the $\sigbcg \- \sigcl$ dynamical scaling relation by tracing the progenitors of clusters and BCGs at $z < 1$. \citet{Sohn22} also derive the dynamical scaling relation between clusters and their BCGs at $z < 1$. They identify the most massive halos with $M_{200} > 10^{14}~\Msun$ in each redshift snapshot and investigate the dynamical properties of clusters and BCGs. This approach yields a direct comparison sample for high redshift observations. 

Here we use IllustrisTNG-300 simulations to explore the redshift evolution of dynamical properties of massive clusters and their BCGs by tracing the progenitors of massive cluster halos ($M_{200} > 10^{14}~\Msun$) at $z = 0$ directly. Based on this approach, we explore the evolution of dynamical properties of clusters and BCGs at $z < 2$. We also investigate the redshift evolution of dynamical scaling relations. The dynamical properties and the dynamical scaling relations we derive from the simulations promise interesting tests for future large spectroscopic surveys covering the full redshift range $z \lesssim 2$. These tests are subtle because of the observational challenge of connecting progenitors and descendants.

We describe the cluster samples we derive from IllustrisTNG simulations in Section \ref{sec:data}. We derive the physical properties of clusters and BCGs. We explore the redshift evolution of scaling relations and trace the redshift evolution of individual systems in Section \ref{sec:results}. We discuss the implications of the dynamical scaling relations along with basic approaches to observational tests based on the prediction from the simulations in Section \ref{sec:discussion}. We conclude in Section \ref{sec:conclusion}. We use the Planck cosmological parameters \citep{Planck16} with $H_{0} = 67.74~\kmsmpc$, $\Omega_{m} = 0.3089$, and $\Omega_{Lambda}$ = 0.6911 throughout the paper.

%========================================
\section{Clusters and BCGs from IllustrisTNG} \label{sec:data}
%========================================

We explore the nature and evolution of scaling relations between the mass and velocity dispersion of clusters and their most massive subhalo (BCG) based on large hydrodynamic simulations. We use the IllustrisTNG simulations that provide a large set of clusters covering the redshift range $z < 2$. We describe the IllustrisTNG simulation in Section \ref{sec:TNG}. We derive the mass and the velocity dispersion of the sample clusters in Section \ref{sec:cl} and those of their brightest cluster galaxies in Section \ref{sec:bcg}. 

\subsection{The IllustrisTNG Simulation}\label{sec:TNG}

IllustrisTNG is a suite of state-of-art magnetohydrodynamic (MHD) simulations of galaxy formation and evolution \citep{Pillepich18b, Springel18, Nelson18, Naiman18, Marinacci18, Nelson19} that succeeds Illustris simulations \citep{Vogelsberger13, Vogelsberger14b}. The large scale of IllustrisTNG allows investigation of galaxy and galaxy system formation in a cosmological context. Here we explore the evolution of massive galaxies (BCGs) in rare massive systems (i.e., galaxy clusters). We complement and extend studies by \citet{Sohn22} based on IllustrisTNG and by \citet{Marini21} based on other simulations.

IllustrisTNG 300-1 covers $\sim 300$ Mpc$^{3}$ comoving cube. TNG300-1 (hereafter TNG300) has the highest resolution among simulations covering this box size: the dark matter particle mass $m_{DM} = 59 \times 10^{6}~\Msun$ and the target gas cell mass $m_{baryon} = 11 \times 10^{6}~\Msun$. The minimum gravitational softening length is 1.48 kpc. 

To identify clusters of galaxies, we use the group catalog provided by the TNG project. The TNG collaboration constructs the group catalog based on a friends-of-friends algorithm applied to all types of particles (including dark matter, stars, gas, and black holes) with a proportional linking length $b = 0.2$. Group catalogs are available for each redshift snapshot. Thus it is possible to explore cluster redshift evolution. 

\begin{figure}
\centering
\includegraphics[scale=0.28]{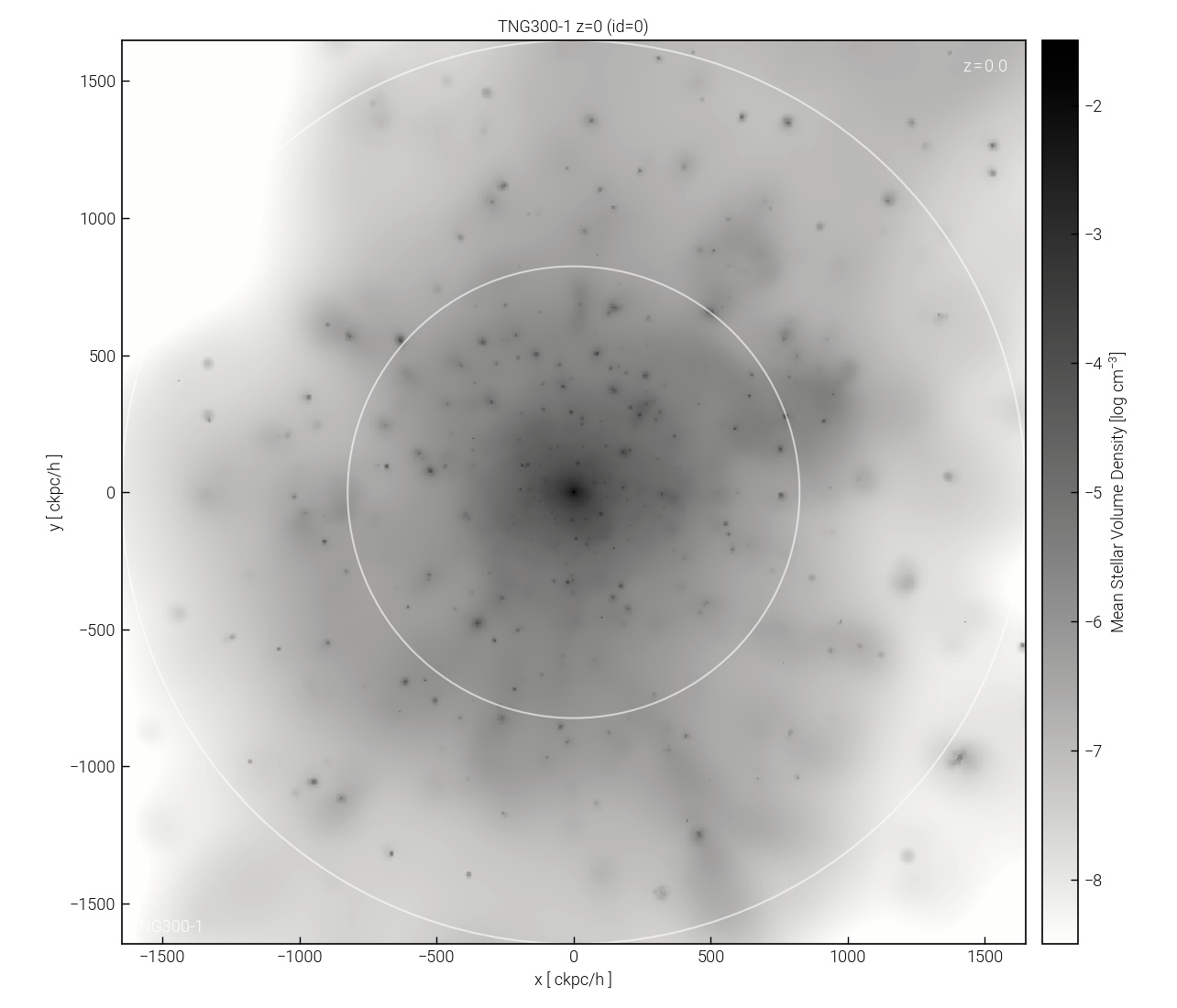}
\caption{Mean density distribution of stellar particles belonging to the most massive cluster in IllustrisTNG-300 $z =0$ snapshot. The darker color indicates higher density. Two circles indicate the $R_{200}$ and $2R_{200}$ of the cluster halo. }
\label{fig:image}
\end{figure}

We select the 280 most massive halos with $M_{200} > 10^{14}~\Msun$ (hereafter clusters) from the $z = 0$ snapshot in TNG300. Here, $M_{200}$ is the characteristic mass enclosed within $R_{200}$, where the mass density is 200 times of the mean density of the universe at a given redshift (in this case, $z = 0$). Figure \ref{fig:image} shows the mean stellar particle density distribution around the most massive cluster halo at $z = 0$. We also select the most massive subhalo within each of these these halos as the Brightest Cluster Galaxy (BCG).

%========================================
%========================================

Next we use the merger trees for identifying the progenitors of both the clusters and their BCGs. The SubLink algorithm \citep{RodriguezGomez15} provides the merger trees. Based on these merger trees, we identify the progenitors of each BCG at $z = 0$ in the relevant higher redshift snapshots with $z \leq 2$. We note that we only trace the main progenitor branch (by applying ``onlyMPB = True"). 

Sometimes a BCG ($\lesssim 10\%$) at lower redshift has a progenitor that is not the most massive subhalo within its higher redshift parent cluster. BCGs can exchange mass during interactions with surrounding subhalos. We include these lower mass BCG progenitors in our analysis. 

There are 10 BCGs without a clear progenitor at higher redshift. Among these, 8 BCGs lack an obvious progenitor in a single redshift snapshot; an additional two BCGs lack obvious progenitors in multiple snapshots at $z > 0.1$ and presumably formed at a very recent epoch. We exclude all of these 10 BCGs and their host clusters from our analysis. Our final simulated cluster sample then includes 270 systems. 

Our approach here complements the approach to the study of clusters and their BCGs in IllustrisTNG by \citet{Sohn22}. Sohn et al. identify all massive clusters with $M_{200} > 10^{14}~\Msun$ within each redshift snapshot and investigate the statistical evolution of the relationship between the velocity dispersions of the clusters and their BCGs. This approach mimics standard observational approaches. \citet{Sohn22} do not identify individual clusters/BCGs with their specific progenitors as we do here. Here, by contrast, we study the evolution of individual clusters and their BCGs by tracking their progenitors back in time. Our approach here highlights evolutionary effects that are not directly observable. However the understanding gained is an important ingredient for motivating and interpreting future observations.

\subsection{Physical Properties of Clusters}\label{sec:cl}

We obtain the characteristic size and mass (i.e., $R_{200}$ and $M_{200}$) of 270 massive cluster halos from the TNG dataset. The cluster mass is the sum of the masses of all of the simulated components: stellar/gas particles, dark matter, and black holes. 

%========================================
\begin{figure}
\centering
\includegraphics[scale=0.18]{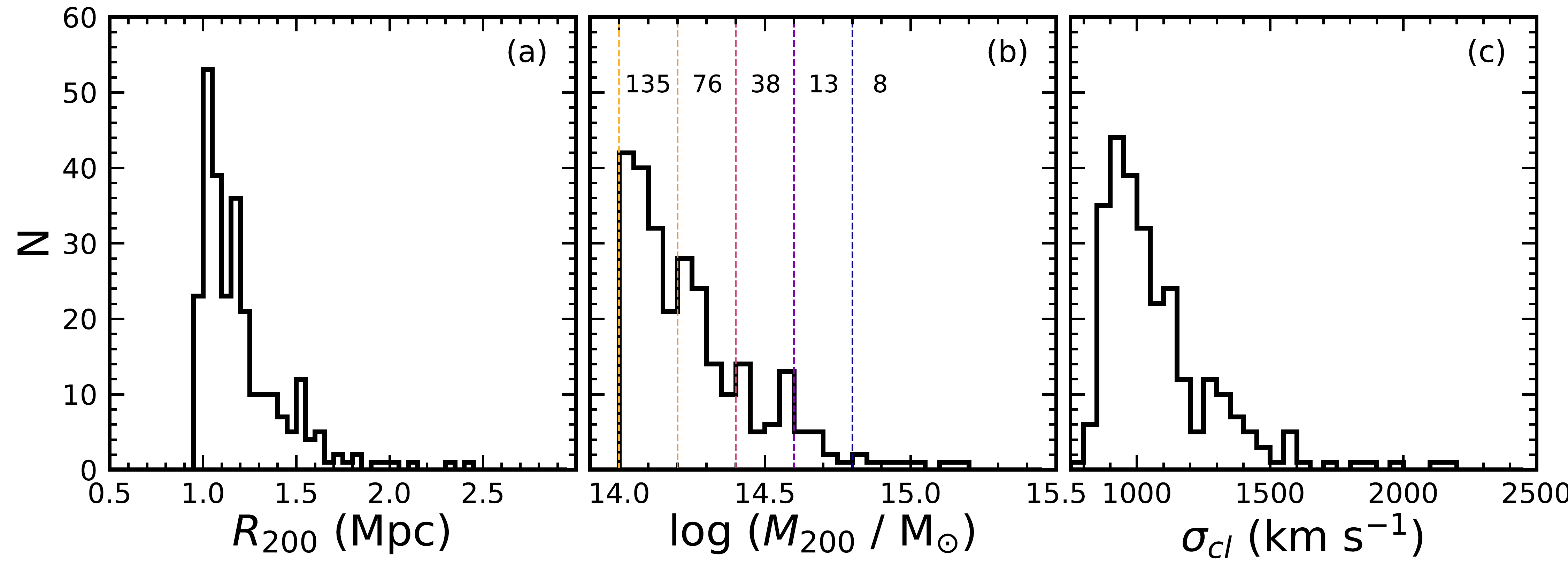}
\caption{(a) $R_{200}$, (b) $M_{200}$, and (c) $\sigma_{cl, 200}$ distributions of 270 massive clusters in TNG300 at $z = 0$. Vertical lines in (b) show the boundaries we use for constructing mass subsamples. }
\label{fig:CL_prop_0}
\end{figure}
%========================================

Figure \ref{fig:CL_prop_0} (a) shows the $R_{200}$ distribution of 270 massive cluster halos at $z = 0$. The median $R_{200}$ for simulated clusters with $M_{200} > 10^{14}~\Msun$ is $1.14 \pm 0.23$ Mpc. Figure \ref{fig:CL_prop_0} (b) displays the $M_{200}$ distribution of the simulated clusters. We trace cluster evolution in five $M_{200}$ bins indicated by vertical lines in Figure \ref{fig:CL_prop_0} (b). Table \ref{tab:mass_sub} lists the mass ranges and the number of clusters in each mass subsample. 

%=================================
%Table \ref{BCG_cat}
%=================================
\begin{deluxetable}{ccc}
\label{tab:mass_sub}
\tablecaption{Number of Clusters in Each Mass Subsample}
\tablecolumns{3}
\small
\tablewidth{0pt}
\tablehead{\multirow{2}{*}{Mass subsample} & \colhead{Mass range} & \multirow{2}{*}{N$_{halo}$} \\
 & [$\log (M_{200} / \Msun)$] & }
\startdata
msub0 & $14.8 - 15.5$ &   8 \\
msub1 & $14.6 - 14.8$ &  13 \\
msub2 & $14.4 - 14.6$ &  38 \\
msub3 & $14.2 - 14.4$ &  76 \\
msub4 & $14.0 - 14.2$ & 135
\enddata 
\end{deluxetable}
%=================================

Figure \ref{fig:CL_prop_0} (c) shows the velocity dispersion distribution of the clusters. We compute the velocity dispersion of member subhalos as a cluster velocity dispersion following \citet{Sohn22}. We select the member subhalos within $R_{cl} < R_{200}$, where $R_{cl}$ is the distance from the cluster center\footnote{$R_{cl} = \sqrt{\Delta X^{2} + \Delta Y^{2} + \Delta Z^{2}}$. $\Delta X, Y, Z$ are the distances between the subhalo and the cluster center along the x-, y-, and z-axes.}. We then compute the cluster velocity dispersion (hereafter $\sigcl$) by summing the velocity dispersions along the axes in quadrature: $\sigcl = \sqrt{(\sigma_{X}^2 + \sigma_{Y}^2 + \sigma_{Z}^2)}$. We use the bi-weight technique \citep{Beers90} to compute the components of the velocity dispersion. 

The cluster velocity dispersion we derive here differs from that measured in \citet{Sohn22}. In \citet{Sohn22}, the cluster velocity dispersion mimics the observed velocity dispersion by computing the line-of-sight velocity dispersion of the subhalos within a cylindrical volume that penetrates the cluster core. In contrast, we compute the 3D velocity dispersion of the subhalos within a spherical volume to elucidate the physical relationship between the cluster and BCG properties in each redshift view. 

%========================================
\begin{figure*}
\centering
\includegraphics[scale=0.37]{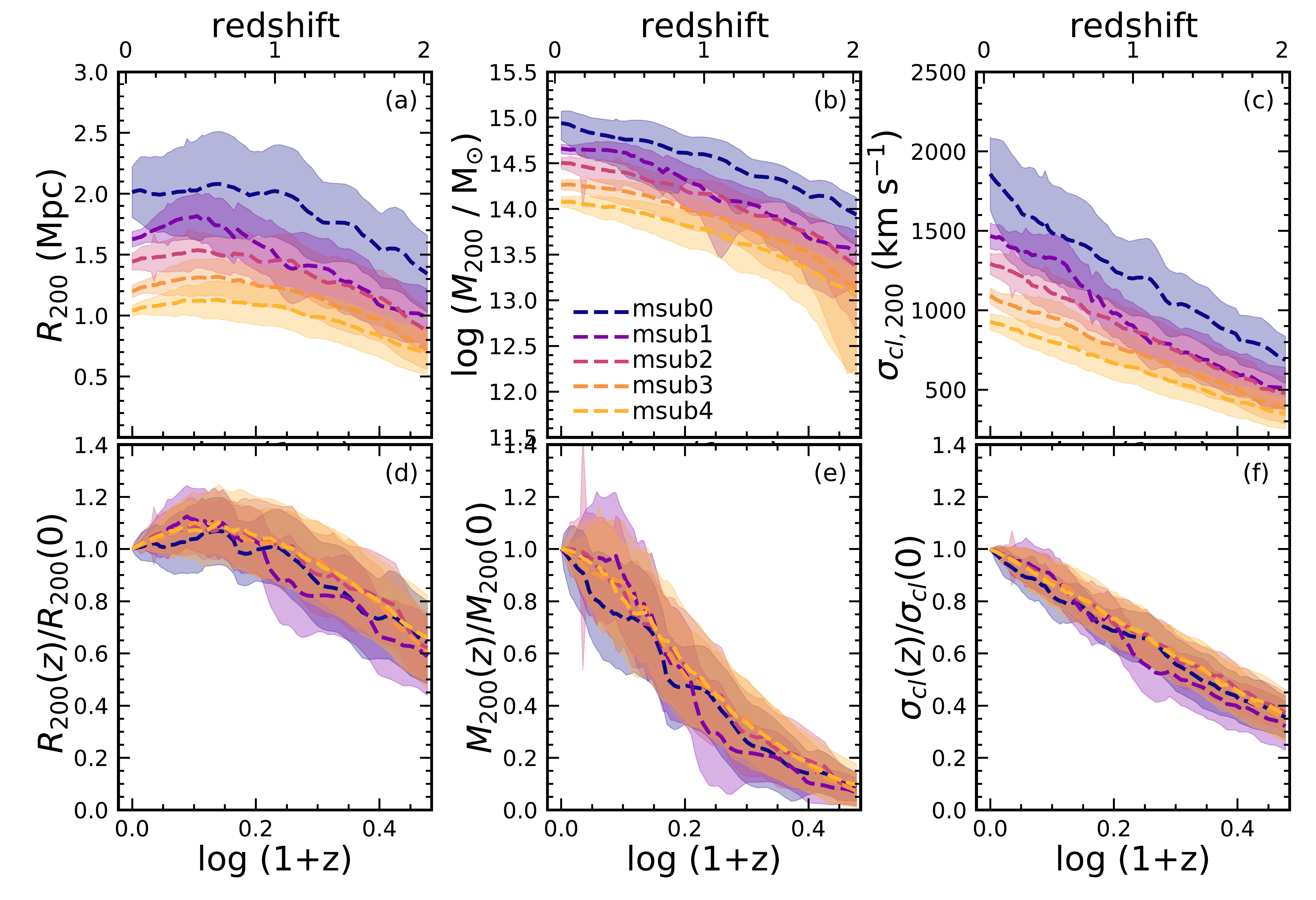}
\caption{Redshift evolution of (a) $R_{200}$, (b) $M_{200}$, and (c) $\sigcl$ of 270 simulated cluster halos. Different colors indicate the mass subsamples (shown in (b)) from Table \ref{tab:mass_sub}. The shaded regions show the $1\sigma$ distribution; the dashed lines indicate the median evolution. (d) - (f) are the same as (a) - (c), but y-axes show the normalized size, mass, and velocity dispersion relative the same quantities measured at $z = 0$. }
\label{fig:CL_evol}
\end{figure*}
%========================================

Figure \ref{fig:CL_evol} shows the evolution of (a) $R_{200}$, (b) $M_{200}$, and (c) $\sigcl$ as a function of redshift for $z < 2$. The dashed lines show the median evolutionary trend. The shaded region shows the $1\sigma$ distribution around the median trend. We indicate the evolution of clusters in five mass subsamples (Table \ref{tab:mass_sub}) with different colors. In general, cluster size, mass, and velocity dispersion increase as universe ages. However, their evolution, particularly the size evolution, fluctuates because cluster halos experience mergers and other dynamical interactions with neighboring halos. During these interactions, components of interacting halos mix thus producing departures from the median relations. Figure \ref{fig:CL_evol} panels (d), (e), and (f) display the evolution of $R_{200}$, $M_{200}$, and $\sigcl$ normalized to the relations at $z = 0$, respectively. The evolutionary trends in the five mass subsamples are similar.

\subsection{Physical Properties of Brightest Cluster Galaxies}\label{sec:bcg}

\begin{figure}
\centering
\includegraphics[scale=0.18]{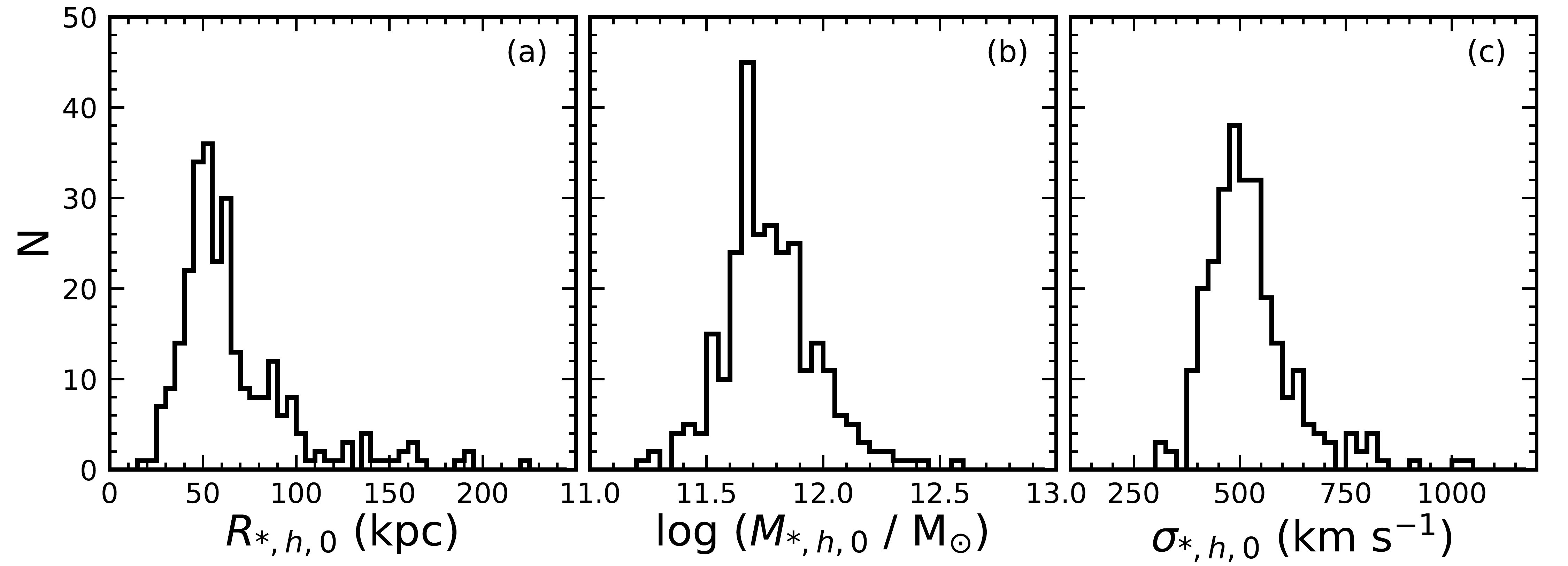} 
\caption{Distributions of (a) half-mass radius ($R_{*, half, 0}$), (b) stellar mass within the half mass radius ($M_{*, h}$), and (c) stellar velocity dispersion of the BCGs ($\sigbcgh$) in 270 massive clusters at $z = 0$.}
\label{fig:BCG_prop_0}
\end{figure}

Next we derive the physical properties of the BCG subhalos, including their sizes, stellar masses, and the velocity dispersions. Figure \ref{fig:BCG_prop_0} (a) and (b) display distributions of the half mass radius ($R_{*, h}$) and the stellar mass within the half mass radius ($M_{*, h}$) of the BCGs at $z = 0$. We obtain the half mass radius and the stellar mass of the BCGs within the half mass radius from the TNG database (i.e., {\it SubhaloMassInHalfRadType{[4]}}). The stellar mass is the sum of the masses of the stellar particles within the half mass radius of the subhalo. 

%========================================
%========================================

We compute the velocity dispersion among the stellar particles of each BCG subhalo. We select the stellar particles within a spherical volume with the half mass radius ($R_{*, h}$). As for the cluster velocity dispersion, we compute the 3D velocity dispersion of the BCG stellar particles based on the quadratic sum of the velocity dispersions within the half mass radius along the three principle axes. Hereafter, we refer to this stellar velocity dispersion as $\sigbcg$. 

The $\sigbcg$ we derive here differs from the BCG velocity dispersion in \citet{Sohn22}. Sohn et al. measure the line-of-sight velocity dispersion of the BCG stellar particles within a cylindrical volume with a 3 kpc physical aperture that penetrates the BCG center. The $\sigbcg$ here is measured within a larger aperture and in a 3D spherical volume. This $\sigbcg$ has greater fidelity for demonstrating the simulated physical relationship between the cluster and BCG properties at different redshifts.

%========================================
\begin{figure*}
\centering
\includegraphics[scale=0.37]{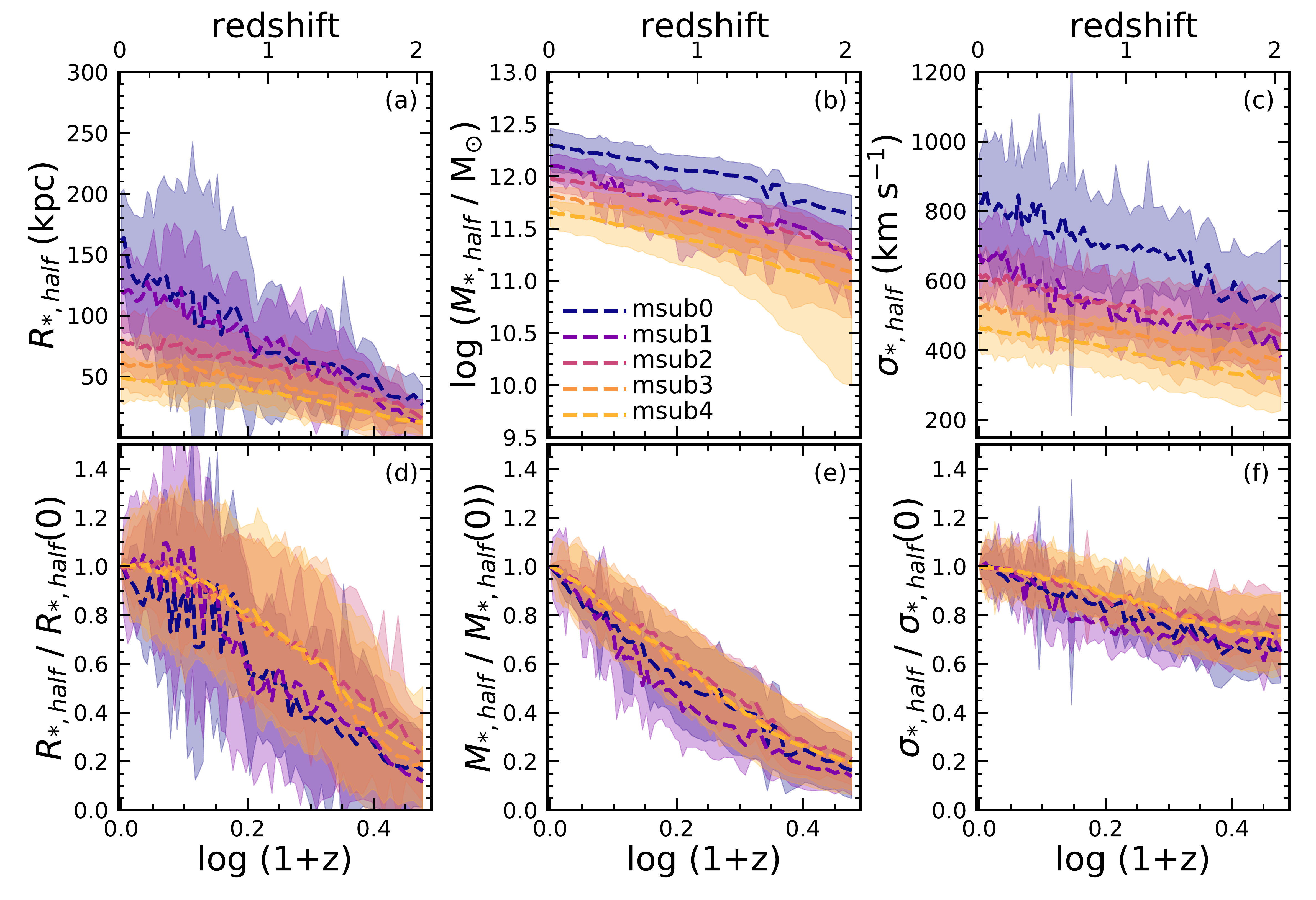}
\caption{Evolution of BCG physical properties including (a) $R_{*, h}$, (b) $\log M_{*, h}$, and (c) $\sigbcg$ as a function of redshift for $z < 2$. The dashed lines indicate the median evolution; the shaded regions show the $1\sigma$ distribution. The lower row displays the normalized evolution of (d) $R_{*, h} / R_{*, h} (z=0)$, (e) $M_{*, h} / M_{*, h} (z=0)$, and (f) $\sigbcg / \sigbcg (z=0)$.}
\label{fig:BCG_evol}
\end{figure*}
%========================================

We trace the redshift evolution of BCG properties for clusters in the five cluster mass subsamples in Table \ref{tab:mass_sub} (Figure \ref{fig:BCG_evol}). The sizes of the BCGs generally increase as the universe ages, but the size evolution fluctuates significantly. These fluctuations result from dynamical interactions between the BCGs and surrounding subhalos. During these interactions, stellar particles of the BCG subhalos can be distributed to larger radii thus inflating the size.

Both the stellar mass and stellar velocity dispersion of BCGs increase as the universe ages. The redshift evolution of these properties also oscillates as a result of interactions between BCGs and surrounding subhalos. 

Nonetheless Figure \ref{fig:BCG_evol} shows that the BCG stellar mass and velocity dispersion growth rates are insensitive to the halo mass of the host cluster. The upper panels of Figure \ref{fig:BCG_evol} show the evolution; the lower panels show the evolution scaled to zero redshift and demonstrate the insensitivity to the cluster halo mass. Both the mass and velocity dispersion growth of the BCGs are slower than the growth of the parent cluster halos.

%========================================
\section{RESULTS} \label{sec:results}
%========================================

We derive the physical properties of clusters and their BCGs from multiple redshift snapshots in IllustrisTNG-300. To highlight the relationship between the BCGs and their host clusters, we explore the redshift evolution of various dynamical scaling relations between clusters and their BCGs (Section \ref{sec:scaling}). We also trace the evolution of individual clusters and BCGs in Section \ref{sec:trace}.  

%========================================
\subsection{Redshift Evolution of Dynamical Scaling Relations} \label{sec:scaling}
%========================================

%========================================
\begin{figure*}
\centering
\includegraphics[scale=0.45]{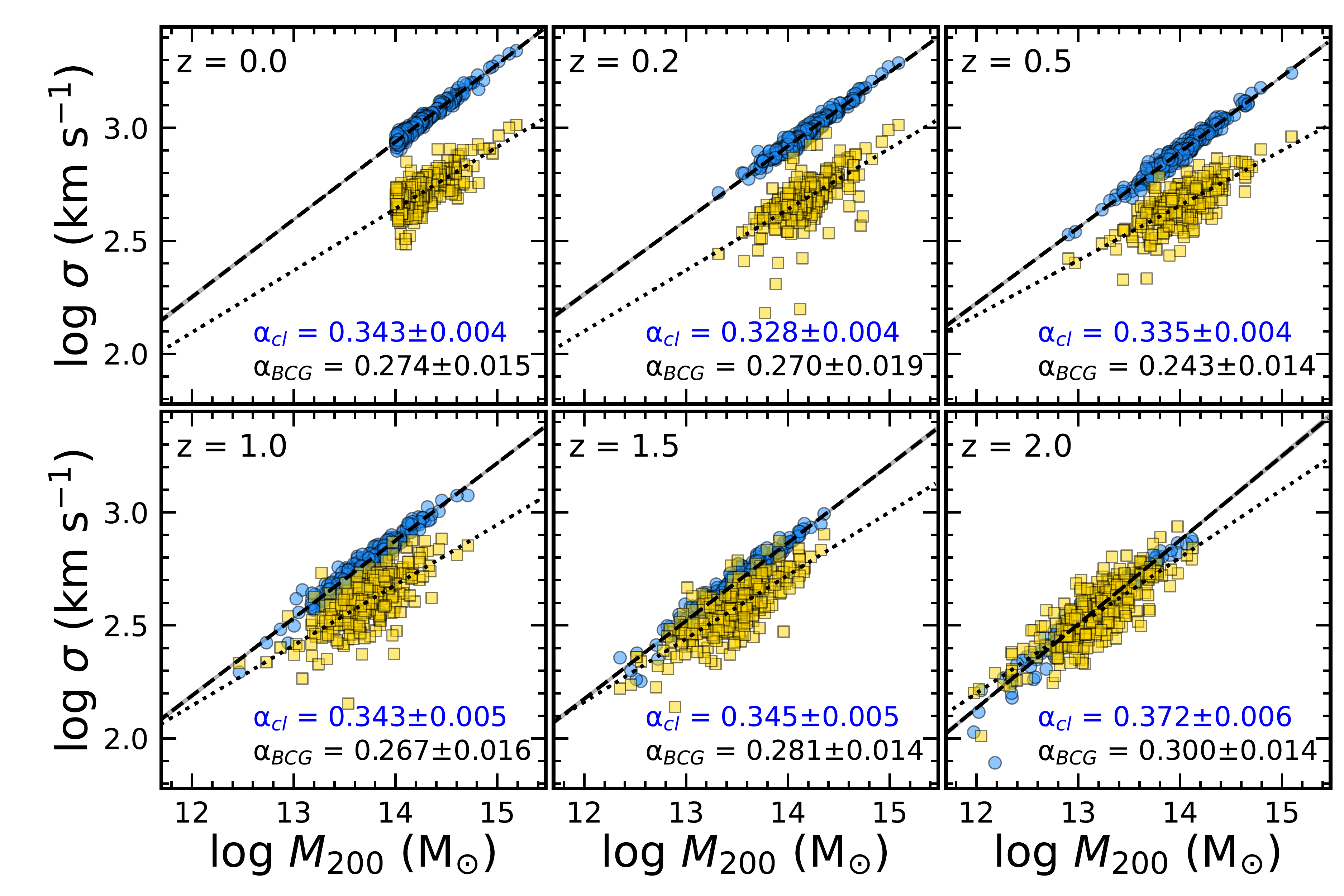}
\caption{Cluster and BCG velocity dispersion as a function of $M_{200}$. Blue circles show $\sigcl$ and yellow squares show $\sigbcg$. Dashed and dotted lines display the best-fit relations; each panel lists the slopes. }
\label{fig:msigma}
\end{figure*}
%========================================

We first investigate the redshift evolution of the $M_{200} \- \sigma$ relation (Figure \ref{fig:msigma}) based on five redshift snapshots (i.e., $z = 0.0, 0.2, 0.5, 1.0, 1.5, 2.0$). In Figure \ref{fig:msigma}, blue circles show $\sigcl$ and yellow squares show $\sigbcg$ as a function of $M_{200}$. We derive the best-fit relation based on the MCMC approach; dashed and dotted lines indicate the best-fit relations. The numbers in each panel indicate the slope of the relation: $\log \sigma = \alpha \log M_{200} + \beta$. 

The $M_{200} \- \sigcl$ relation is generally steeper than the $M_{200} \- \sigbcg$ relation at $z < 2$. The slope of the $M_{200} \- \sigcl$ relation is $\sim 0.33$, consistent with previous studies \citep{Marini21, Sohn22}. The normalization of the relation is slightly higher than that in \citet{Sohn22} because they use the line-of-sight cluster velocity dispersion (instead of the 3D $\sigcl$) for direct comparison with the observations. In contrast, the $M_{200} \- \sigbcg$ relation is significantly shallower than the $M_{200} \- \sigcl$ relation; the typical slope at $z < 2$ is $\sim 0.27 \pm 0.02$. Because we also use the 3D $\sigbcg$ rather than line-of-sight BCG stellar velocity dispersion, the normalization of the relation is higher than in \citet{Sohn22}. 

\citet{Sohn22} investigated the $M_{200} \- \sigma$ relation based on the observed cluster sample, HeCS-omnibus. They also showed that the $M_{200} \- \sigbcg$ relation is significantly shallower than the $M_{200} \- \sigcl$ relation at $z < 0.15$. Furthermore, the typical slopes of the observed relations are consistent with the relations we derived from the simulations.

It is interesting that the difference between $\sigcl$ and $\sigbcg$ is smaller at higher redshift. In fact, the $M_{200} \- \sigcl$ and $M_{200} \- \sigbcg$ relations overlap at $z = 2$. Thus $\sigcl$ increases significantly faster as the universe ages than $\sigbcg$ does.  

\begin{figure*}
\centering
\includegraphics[scale=0.45]{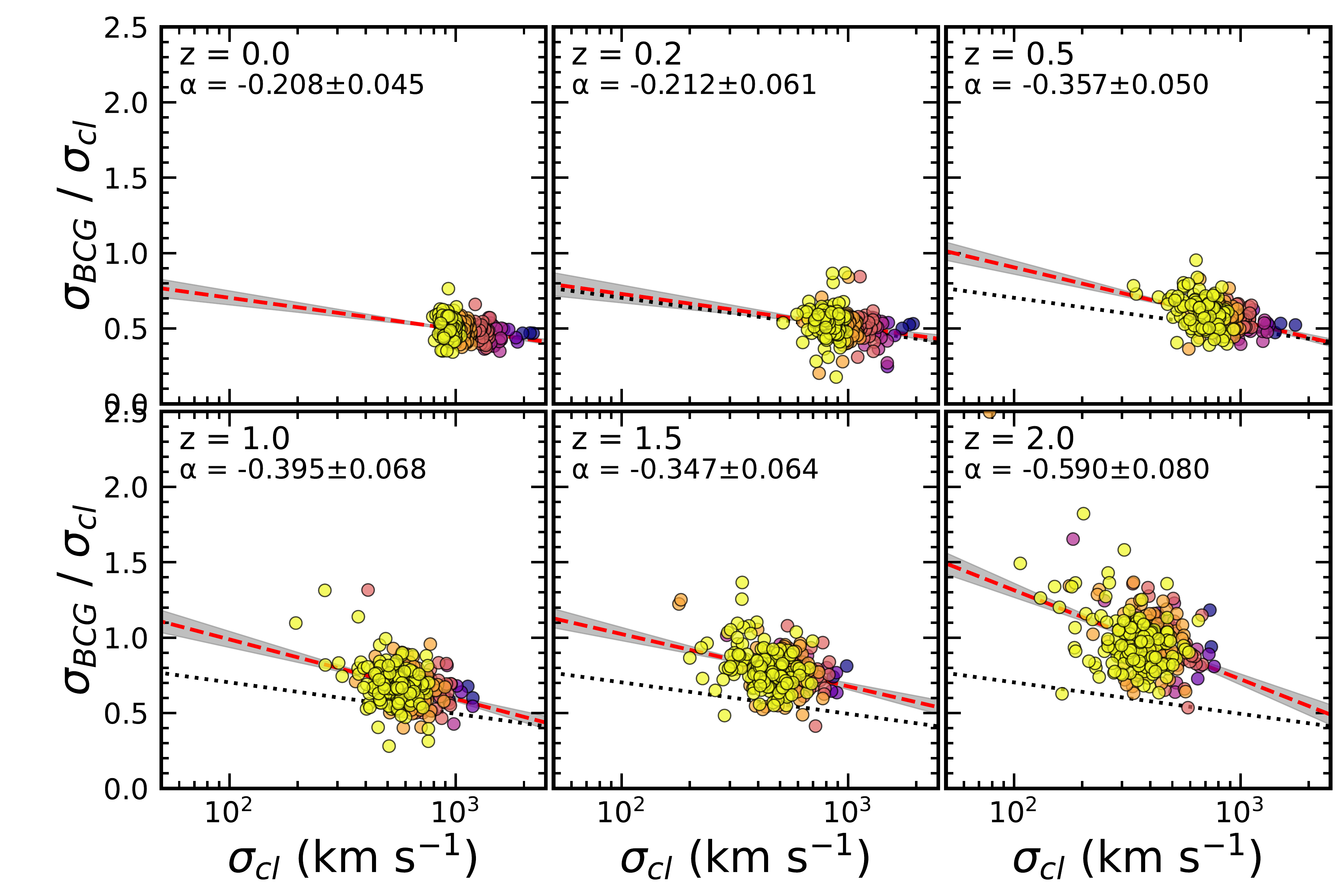}
\caption{The $\sigbcg / \sigcl$ ratio as a function of $\sigcl$. Circles show individual clusters color coded by their mass at $z = 0$; a darker color indicates a more massive cluster at the current epoch. Red dashed lines and the shaded regions mark the best-fit and the scatter derived from the MCMC approach. For comparison, black dotted lines display the relation at $z = 0$.}
\label{fig:sigma_ratio_redshift}
\end{figure*}

Figure \ref{fig:sigma_ratio_redshift} shows the $(\sigbcg / \sigcl) \- \sigcl$ relation of simulated clusters from six redshift snapshots. In general, the $(\sigbcg / \sigcl)$ ratio decreases as a function of $\sigcl$. The $(\sigbcg / \sigcl) \- \sigcl$ relation becomes significantly steeper at higher redshift. In other words, the $\sigbcg$ is comparable with $\sigcl$ at high redshift.

The sample completeness is an important issue when we trace the evolution of galaxy systems over a wide redshift range. In \citet{Sohn22}, we select a mass complete sample at different redshifts to study the evolution of scaling relations between $\sigcl$ and $\sigbcg$. In contrast, our focus here is tracing the evolution of the progenitors of clusters in the mass complete sample at $z = 0$. At high redshifts, the progenitors of the currently most massive systems may not constitute a mass-complete because of fluctuations in the mass evolution but they still trace the scaling relation defined by the mass complete sample of \citet{Sohn22}.

%========================================
%========================================

%========================================
\subsection{Tracing the Evolution of Simulated Clusters} \label{sec:trace}
%========================================

We next investigate the evolution of $\sigcl$ and $\sigbcg$ for individual clusters and their BCGs at $z < 2$. This approach traces the evolution of clusters and the BCGs as a function of time. We trace the dependence of cluster and BCG evolution on the cluster halo mass. 

%========================================
\begin{figure*}
\centering
\includegraphics[scale=0.28]{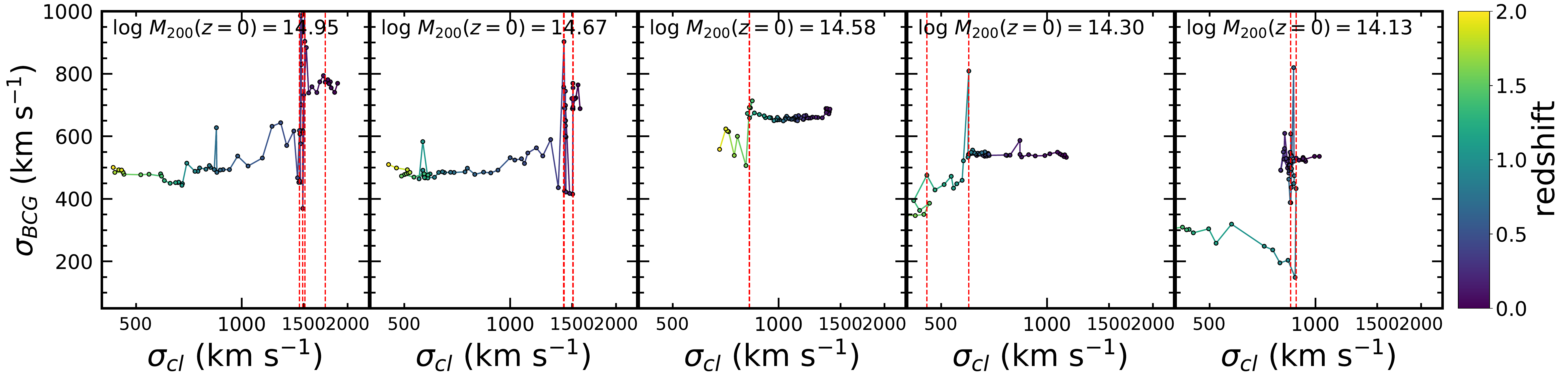}
\caption{Redshift evolution of $\sigbcg \- \sigcl$ relation for five randomly selected simulated clusters. In each panel, clusters evolve from $z = 2$ to $z = 0$ (from left to right). Spikes at some redshifts reflect a change in $\sigbcg$ resulting from dynamical interaction with neighboring galaxies. Red vertical lines indicate the occurrence of a major merger. }
\label{fig:trace_ind}
\end{figure*}
%========================================

Figure \ref{fig:trace_ind} shows the evolution of the $\sigbcg \- \sigcl$ relation for five simulated clusters  randomly selected from each of the five mass subsamples (Table \ref{tab:mass_sub}). The systems generally move from left to right as $\sigcl$ increases for $z < 2$. The $\sigbcg$ of individual systems generally slowly increases as the universe ages. 

The $\sigbcg \- \sigcl$ relations show a few spikes during the evolution. Abrupt changes in $\sigbcg$ cause these spikes. The BCG velocity dispersion, $\sigbcg$, increases dramatically just after interactions with a neighboring galaxy; subsequently the dispersion settles down. Vertical lines in Figure \ref{fig:trace_ind} mark the redshift epoch when these major mergers occur. Here, we define a major merger as a dynamical interaction where the mass ratio between the BCG subhalo and the merging subhalo is less than 5:1.

%========================================
\begin{figure}
\centering
\includegraphics[scale=0.26]{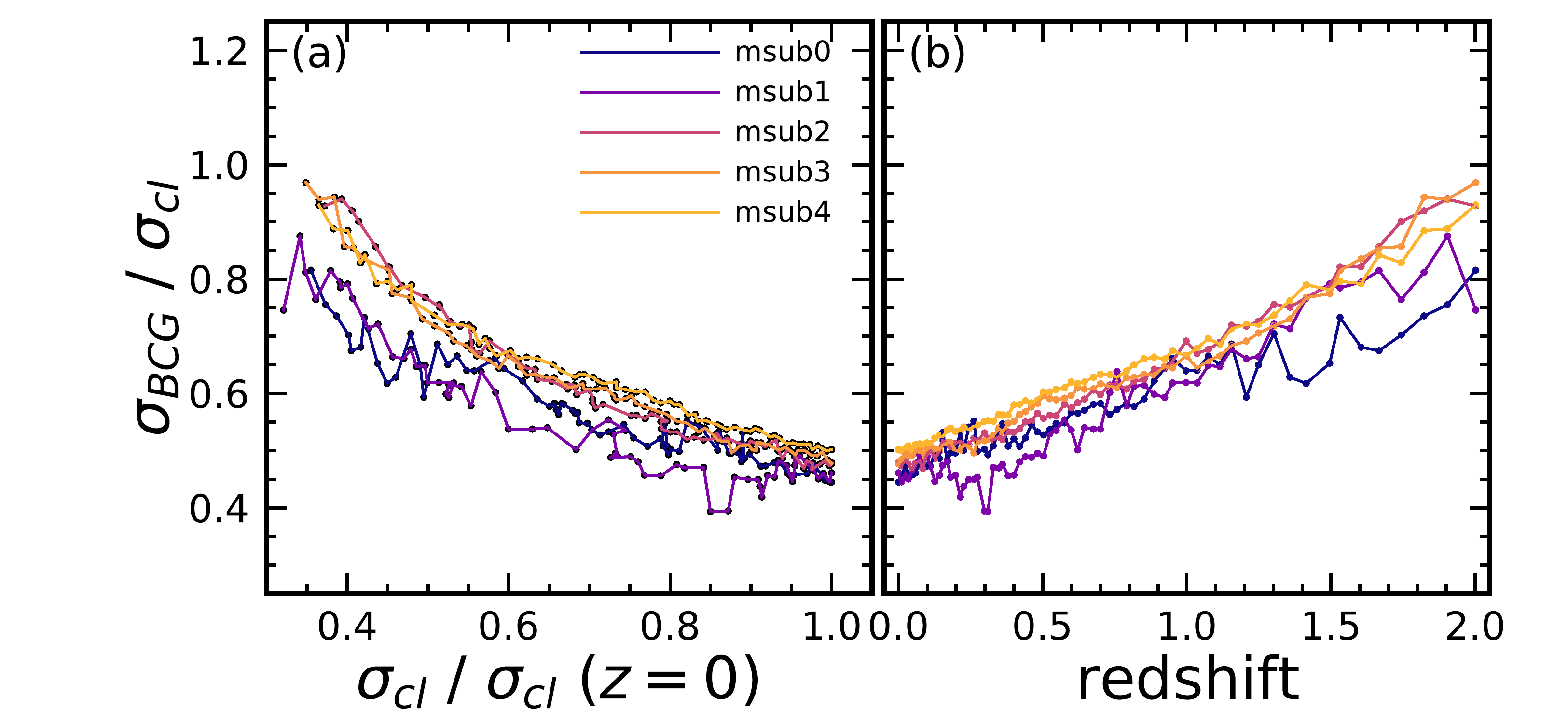}
\caption{(a) Median distribution of $\sigbcg / \sigcl$ as a function of median $\sigcl$ for clusters in the five mass subsamples; different colors indicate the mass subsamples. Individual points are derived from the redshift snapshots for $z < 2$. From left to right, the systems evolve from $z = 2$ to $z = 0$. For direct comparison, we normalize the x-axis ($\sigcl$) by the $\sigcl$ at $z = 0$. (b) Median $\sigbcg / \sigcl$ of the clusters in mass subsamples as a function of redshift. }
\label{fig:trace_msub}
\end{figure}
%========================================

Figure \ref{fig:trace_msub} (a) displays the median $(\sigbcg/\sigcl)$ as a function of $\sigcl$ for the clusters in the five mass subsamples. We derive these parameters to investigate the redshift evolution of the $(\sigbcg / \sigcl) \- \sigcl$ relation. We also normalize the x-axis in Figure \ref{fig:trace_msub} (a) by the median $\sigcl$ of the clusters at $z = 0$. Figure \ref{fig:trace_msub} is thus a direct comparison of the evolution of scaling relations derived from each of the mass subsamples. 

Interestingly, the trends of the $(\sigbcg / \sigcl) \- \sigcl$ evolution are remarkably similar for the five mass subsamples. The relations derived agree within the $1\sigma$ standard deviation (we omit the range from the Figure for the clarity). The slope of the relations steepens at $\sigcl / \sigcl (z=0) < 0.4$. This change in slope coincides with the trends in the redshift evolution of $\sigbcg$ and $\sigcl$. Figure \ref{fig:CL_evol} shows that $\sigcl$ increases steadily for $z < 2$. In contrast, the increase in $\sigbcg$ slows at $z < 1$; $\sigbcg$ not evolve significantly at $z < 0.4$.

Figure \ref{fig:trace_msub} (b) shows the median $(\sigbcg / \sigcl)$ as a function of redshift. In all subsamples, $(\sigbcg / \sigcl)$ decreases continuously at $z < 2$. Furthermore, the slopes of all of the relations are remarkably consistent. This consistency indicates that the increase $\sigcl$ and $\sigbcg$ is insensitive to cluster halo mass for $M_{200} > 10^{14}~\Msun$. 

%========================================
\section{Discussion} \label{sec:discussion}
%========================================

Based on 67 IllustrisTNG-300 snapshots covering $z < 2$, we derive physical properties of the clusters and their BCGs and investigate the redshift evolution of various dynamical scaling relations. 

We first discuss the impact of aperture effects on the BCG velocity dispersion (Section \ref{sec:aperture}). We then compare the dynamical scaling relations we derive with scaling relations based on cluster mass and BCG stellar mass in Section \ref{sec:comparison}. Finally, we discuss tests of the simulated results with current and future spectroscopic surveys (Section \ref{sec:obs}).

\subsection{Aperture Effect and $\sigbcg$} \label{sec:aperture}

In previous sections we used the BCG stellar velocity dispersion, $\sigbcg$, within $R_{*, h}$. This velocity dispersion is tightly correlated with the stellar mass within $R_{*, h}$ throughout the redshift range we explore. Thus, $\sigbcg$ is a mass proxy that complements the stellar mass. Used together the two proxies can be a foundation for limiting systematic error in BCG mass determination.

Observations of $\sigbcg$ are, however, not completely straightforward. Observational determination of the half-mass radius is challenging. Measurement of the half-light radius may be affected by superposition of other objects and by failure to account for the extended BCG halo. Furthermore, the half-light radius is not identical to the half-mass radius. 

The half-mass radius of BCGs derived from the simulations are typically very large. Figure \ref{fig:BCG_prop_0} shows the $R_{*, h}$ distribution; the $R_{*, h}$ values range from 20 to 200 kpc at $z = 0$ corresponding to 16 to 160 arcmin. 

Measurements of the BCG velocity dispersions are often based on fiber spectrographs (e.g., SDSS/BOSS or MMT/Hectospec) that usually cover only the core region of the BCG \citep{Sohn17a, Sohn20} that is much smaller $R_{*, h}$. Thus a significant aperture correction is required to convert from the observations to the larger half-mass radii.

For direct comparison with the observations, \citet{Sohn22} derive the dynamical scaling relations from TNG based on the BCG velocity dispersion measured within 3 kpc at $z = 0$. We note that the BCG velocity dispersion ($\sigbcgc$) used in \citet{Sohn22} differs from the BCG velocity dispersion ($\sigbcgh$) we use here. The scaling relations in \citet{Sohn22} show generally similar trends with the relation we derive here. For example, $\sigma_{*, {\rm BCG, 3~kpc}}$ and $\sigma_{*, {\rm BCG, R_{*, h}}}$ are proportional to $\sigcl$. Both $(\sigma_{*, {\rm BCG, 3~kpc}} / \sigcl)$ and $(\sigma_{*, {\rm BCG, R_{*, h}}} / \sigcl)$ decrease as a function of $\sigcl$. In other words, the behavior of the dynamical scaling relations is relatively insensitive to the aperture size. 

\begin{figure*}
\centering
\includegraphics[scale=0.34]{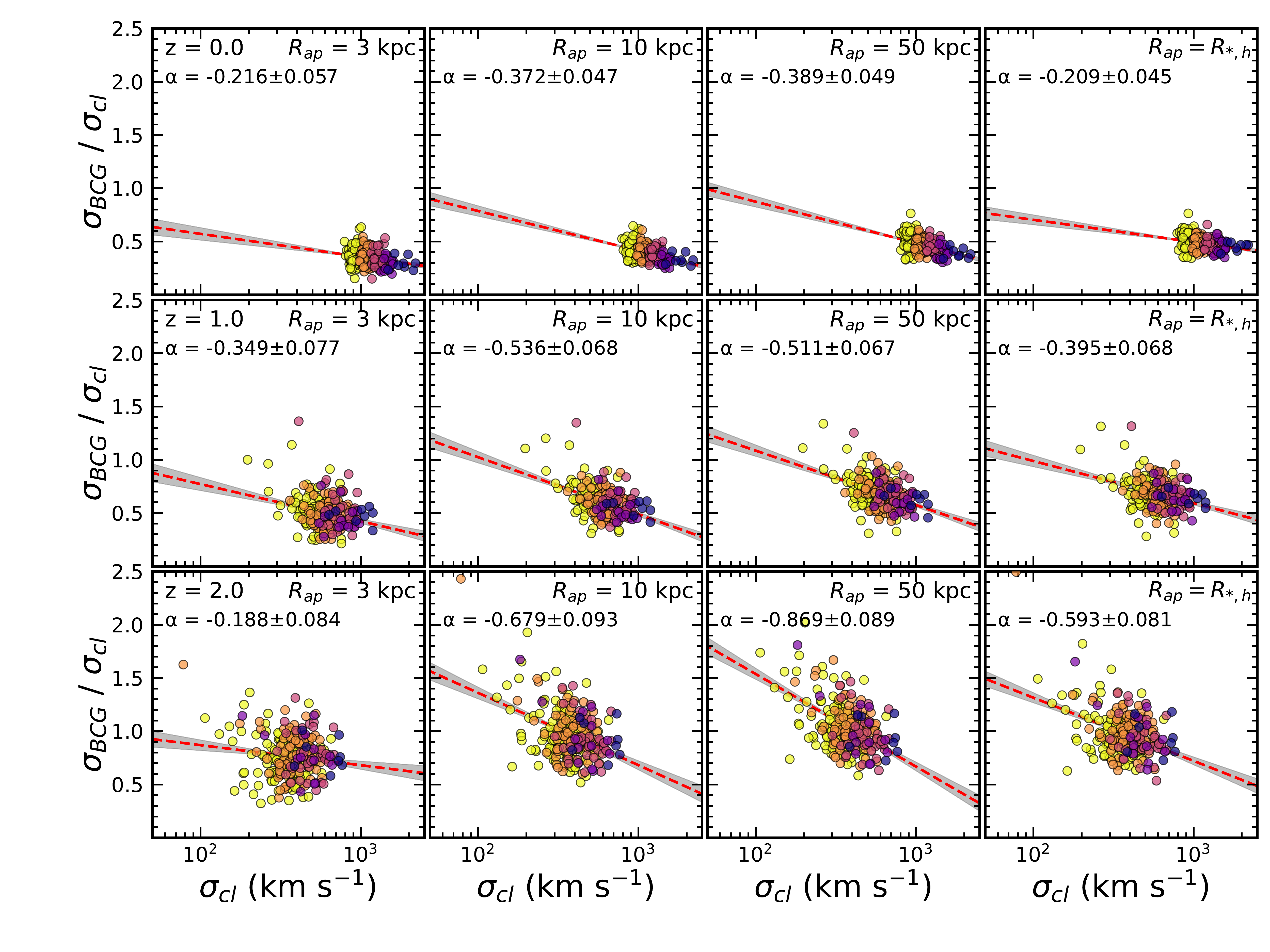}
\caption{The $(\sigbcg / \sigcl) \- \sigcl$ relation for clusters in three redshift snapshots ($z = 0.0, 1.0$, and 2.0 from top to bottom). We show the BCG stellar velocity dispersion measured within 3 kpc, 10 kpc, 50 kpc, and the half-mass radius (from left to right). }
\label{fig:aperture}
\end{figure*}

We next explore the impact of the aperture size on the dynamical scaling relations in more detail. Figure \ref{fig:aperture} displays the $(\sigbcg / \sigcl) \- \sigcl$ relations for systems in three redshift snapshots (i.e., $z = 0.0, 1.0,$ and 2.0). We derive the relations based on the BCG stellar velocity dispersion measured within 3 kpc, 10 kpc, 50 kpc, and $R_{*, h}$ apertures from top to bottom. 

The $(\sigbcg / \sigcl) \- \sigcl$ relations have negative slopes regardless of the BCG velocity dispersion aperture. Red dashed lines show the best-fit relations. The decreasing $(\sigbcg / \sigcl)$ as a function of $\sigcl$ is insensitive to the aperture size used to measure the BCG velocity dispersion.

In general, the $(\sigbcg / \sigcl) \- \sigcl$ relations are steeper at higher redshift regardless of the aperture size. The larger scatter at high redshift indicates that the higher redshift BCGs are usually unrelaxed. \citet{Sohn22} demonstrate that the BCG subhalos at higher redshift ($z \sim 1$) indeed show a more disturbed distribution of the line-of-sight velocities of stellar particles compared to the analogous distributions for their counterparts at low redshift. Indeed, the BCG subhalos at higher redshift we explore generally have complex velocity distributions that drive the larger scatter in the scaling relations. 

The slopes of the $(\sigma_{BCG}/\sigcl) \- \sigcl$ relations are relatively shallower when we use $\sigma_{BCG}$ measured within 3 kpc and $R_{*, h}$ compared to those measured within 10 and 30 kpc. This variation is related to the velocity dispersion profiles. \citet{Bose21} display velocity dispersion profile of halos as a function of projected radius. The stellar velocity dispersion increases as a function of projected radius up to 10 - 100 kpc (depending on the halo mass) and decreases at larger projected radius. Thus, the velocity dispersions measured within 3 kpc and $R_{*, h}$ are generally comparable; the resulting dynamical scaling relations then have similar slopes. 

We note that the scaling relation based on the 3 kpc velocity dispersion at $z = 2$ has a shallower slope and large scatter. Because the stellar particles in the core region of the high redshift BCGs are unrelaxed, the BCG velocity dispersions have a large scatter, resulting in a weaker correlation. \citet{Sohn22} display the phase-space diagrams of the high redshift BCGs (their Figure 12) that show more a disturbed velocity distribution of the stellar particles in the central region compared to their counterpart at low redshift. In other words, using a large aperture for measuring the high redshift BCGs yield more reliable $\sigbcg$ measurement, insensitive to the unrelaxed dynamics of the BCG core region. 

We note that the stellar velocity dispersion within the core region (e.g., $< 4.5$ kpc, 2.8 times the softening length scale, \citealp{Campbell17}) may be underestimated. In this core region, scattering interactions among the particles are softened significantly leading to the probable underestimation of the dispersion.

%========================================
%========================================

%========================================
\subsection{Stellar Mass and Velocity Dispersion as BCG Mass Proxies}\label{sec:comparison}
%========================================

Studying the co-evolution of clusters and their BCGs provides tests of both structure and massive galaxy formation models. In initial explorations of BCG properties and evolution, the luminosities of BCGs were the basis for estimating the BCG mass. For example, \citet{Lin04} showed that more massive clusters host brighter (i.e., more massive) BCGs. More recently, the stellar mass to halo mass relations as a key to BCG evolution have been explored in observations and simulations (e.g., \citealp{Kravtsov18, Erfanianfar19}). The ratio between BCG stellar mass and cluster halo mass decreases as a function of cluster mass at $M > 10^{13}~\Msun$. 

\citet{Sohn20} suggest that the $\sigbcg \- \sigcl$ relation provides an independent observational test for studying the relation between cluster and BCG masses (see also \citealp{Sohn21}). They used $\sigbcg$ and $\sigcl$ as mass proxies for the BCG subhalo and cluster halo masses. 

Many studies demonstrate that $\sigcl$ is a good cluster mass proxy based on the $\sigcl \- M_{200}$ relation. In contrast, $\sigbcg \- M_{200}$ or $\sigbcg \- M_{total, BCG}$ relations are not widely discussed. In particular, the redshift evolution of these relations is not well known. Exploring these relations based on simulations provides an independent assay of the use of $\sigbcg$ as a BCG mass tracer. This theoretical test provides guidance for using $\sigbcg$ from future spectroscopy, particularly of high redshift systems, as a window on the evolution of the most massive galaxies in the universe.

\begin{figure}
\centering
\includegraphics[scale=0.23]{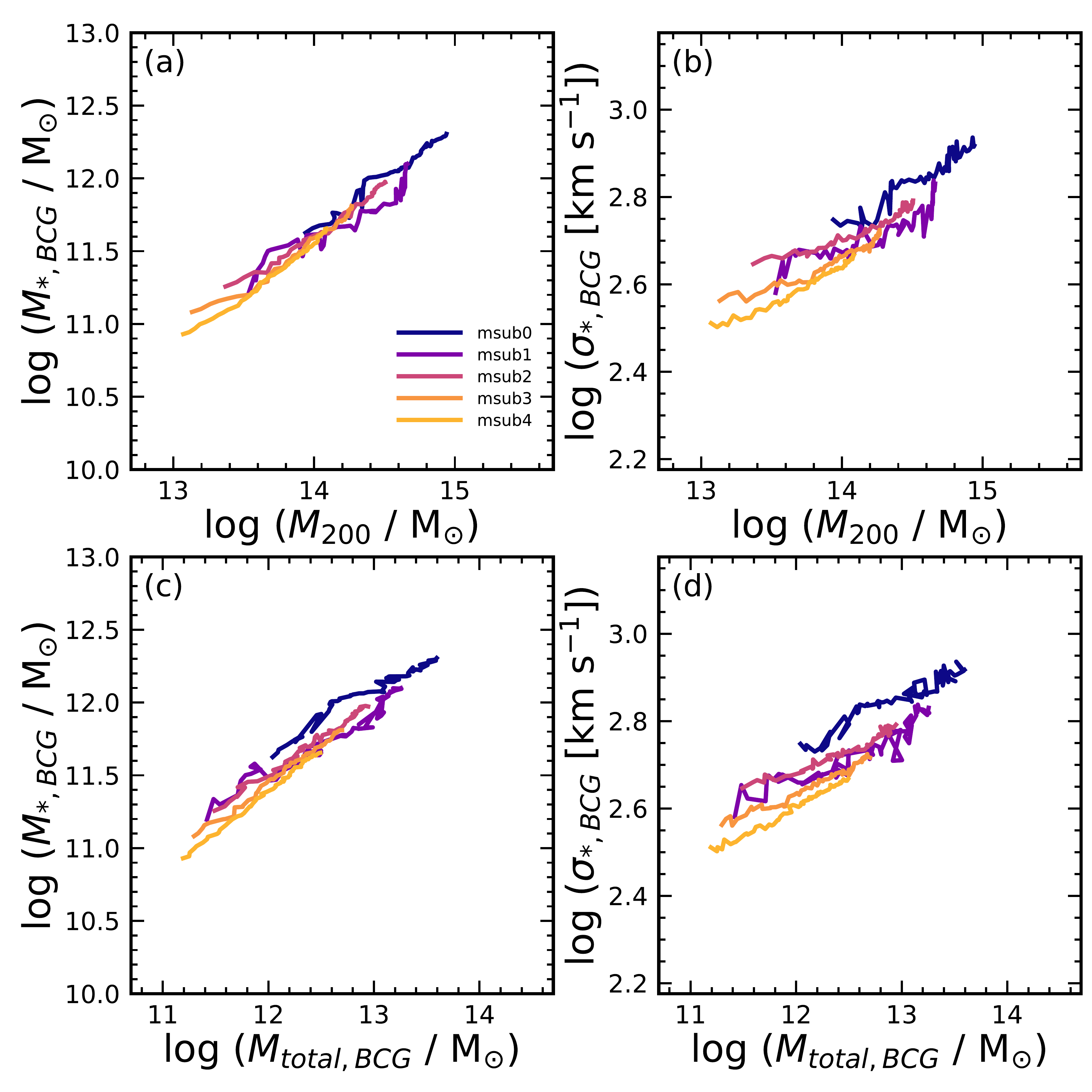}
\caption{Redshift evolution of the (a) BCG stellar mass ($M_{*, BCG}$) and (b) BCG stellar velocity dispersion ($\sigbcg$) to cluster mass ($M_{200}$) relations. Lines connect the median values computed for each redshift snapshot. Different colors indicate the relations derived for each mass subsample. Redshift evolution of (c) M$_{*, {\rm BCG}}$ and (d) $\sigma_{*, {\rm BCG}}$ as a function of BCG total mass. Line colors are the same as in (a) and (b). }
\label{fig:bcg_sigma}
\end{figure}

Figure \ref{fig:bcg_sigma} (a) displays the redshift evolution of the BCG subhalo stellar mass and cluster halo mass relation. We derive the relation based on systems in the usual five mass subsamples (separated with colors). We then identify their progenitors in the higher redshift snapshots. In each redshift snapshot, we compute the median cluster mass and the BCG subhalo mass. The lines show how the median $M_{*, BCG}$ and the median $M_{200}$ change with time; both the median $M_{*, BCG}$ and $M_{200}$ increase as the universe ages (moving from bottom left to upper right). Although the normalization changes, the slope of the redshift evolution for the mass subsamples is similar in all cases. This result suggests that the relative mass growth rate of BCGs and clusters are independent of the host cluster halo mass. 

Figure \ref{fig:bcg_sigma} (b) shows the redshift evolution of the $\sigbcg \- M_{200}$ relation. The general trends are essentially identical to those for the $M_{*, BCG} \- M_{200}$ relations. The median $\sigbcg$ increases monotonically as a function of $M_{200}$. The slopes of the scaling relations are similar in all of the mass subsamples. In other words, the BCG stellar velocity dispersion is a mass proxy that is interchangeable with and complementary to the stellar mass estimates. 

We also investigate the relations between BCG stellar mass and stellar velocity dispersion and the total mass of BCGs (Figure \ref{fig:bcg_sigma} (c) and (d)). Here, the total mass indicates that the sum of dark matter, gas, and stellar particle mass belonging to the BCG subhalo within the half mass radius. Both the $M_{*, BCG} \- M_{total, BCG}$ and $\sigma_{*, BCG} - M_{total, BCG}$ relations show similar redshift evolution. In other words, tracing the $\sigma_{*, BCG} \- M_{total, BCG}$ relation is an independent probe of the co-evolution of clusters and their BCGs. 

%========================================
%========================================

%========================================
\subsection{Observational Tests} \label{sec:obs}
%========================================

We investigate the dynamical properties of clusters and their BCGs available from future spectroscopic surveys reaching $z \lesssim 2$. These dynamical properties can be used for tests of the model developed here for coordinated BCG and cluster evolution. Here, we compare the evolution we trace based on IllustrisTNG with currently available observations and we suggest some future observations. 

One of the distinctive predictions from TNG300 simulation is that BCG and cluster velocity dispersions at higher redshift are comparable. Figure \ref{fig:msigma} and Figure \ref{fig:sigma_ratio_redshift} show that $\sigbcg$ is comparable with $\sigcl$ at $z \gtrsim 1$. At lower redshift $\sigcl$ generally exceeds $\sigbcg$. Future IFU observations will provide the velocity dispersion of the cluster members along with the spatially resolved BCG stellar velocity dispersion for high redshift clusters. A mass complete sample of tens of clusters covering the range $0.8 < z < 1$ would provide a direct test of the $\sigbcg \simeq \sigcl$ when the clusters are identified with their probable descendants. 

\begin{figure*}
\centering
\includegraphics[scale=0.28]{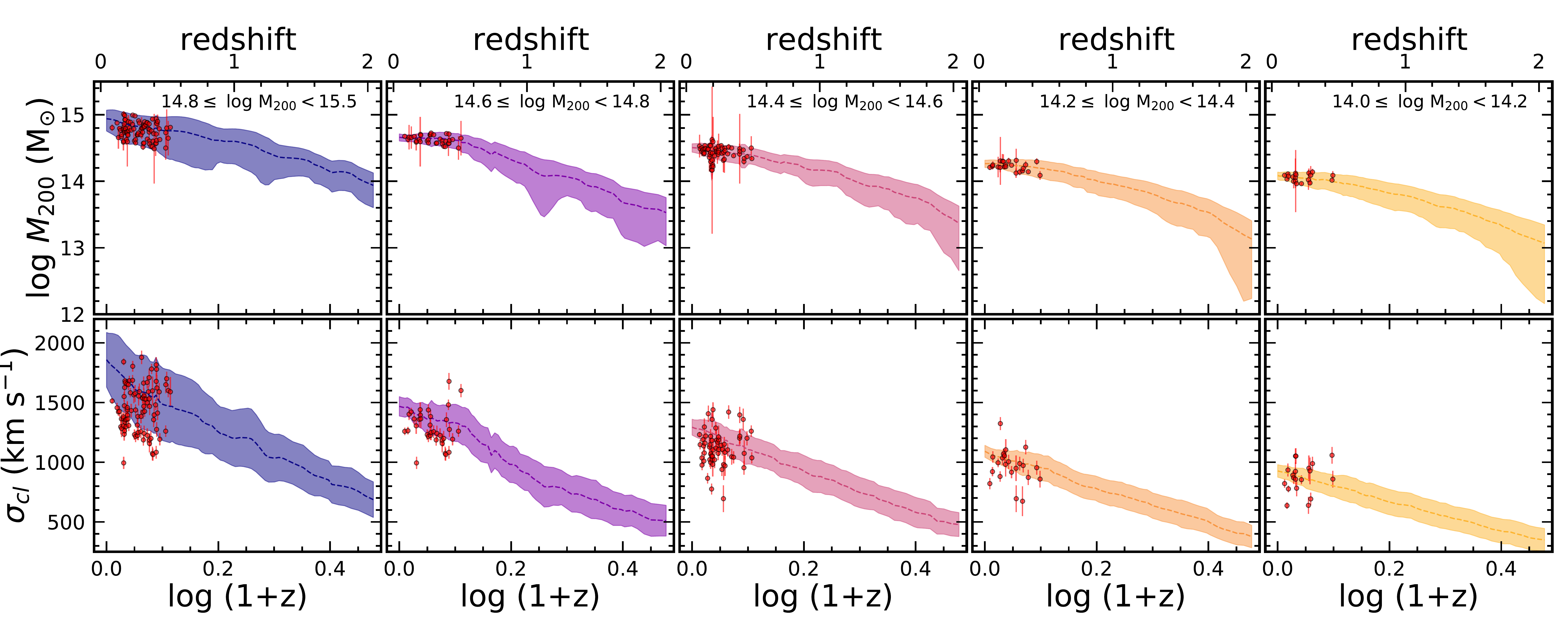}
\caption{$M_{200}$ (upper panels) and $\sigcl$ (lower panels) for simulated clusters (the dashed line and the shaded region) in five zero redshift mass subsamples as a function of redshift. Red circles show the observed HeCS-omnibus clusters. }
\label{fig:cl_obs}
\end{figure*}

Figure \ref{fig:cl_obs} shows the $M_{200}$ and $\sigcl$ evolution as a function of redshift and provides further testable insights. We derive the evolution for clusters in five mass subsamples. Figure \ref{fig:CL_evol} shows that both $M_{200}$ and $\sigcl$ increase as the universe ages. 

We compare the redshift evolution of $M_{200}$ and $\sigbcg$ derived from simulations with observations by making a statistical connection between the observed clusters and the zero redshift mass of their descendant. 

We use the observed HeCS-omnibus sample, a large spectroscopic compilation of 227 clusters at $z < 0.27$ \citep{Rines06, Rines13, Rines16, Rines18, Sohn20}. The HeCS-omnibus clusters typically contain $\sim 180$ spectroscopically identified members. The dense spectroscopy provides a dynamical mass based on the caustic technique \citep{Diaferio97, Diaferio99, Serra13} along with the cluster velocity dispersion \citep{Sohn20}. The uncertainly in the observed cluster mass is generally smaller than the 1$\sigma$ range in the masses of clusters in each of the redshift bins that trace the cluster evolution.

We identify HeCS-omnibus clusters with the ancestors of clusters in the zero redshift bins of Figure \ref{fig:cl_obs} based on the observed caustic masses. At the redshift of each HeCS-omnibus cluster, we compute the $1\sigma$ boundary for the cluster ancestors (progenitors) in each mass subsample. We then place the observed cluster in all bins that includes its caustic mass. A HeCS-Omnibus cluster may appears in one (136) to two (40) or three (3) simulation test bins. 

The HeCS-Omnibus sample includes the most massive clusters in its redshift range. Thus among the 227 HeCS-omnibus clusters, 48 systems are excluded because their masses exceed the mass range of the simulated progenitors. In other words, in spite the size of IllustrisTNG, the simulations are not large enough to include the most massive systems observed at zero redshift. 

In Figure \ref{fig:cl_obs}, red points mark the HeCS-omnibus clusters in each bin. We select the observational comparison sample based on mass: the observed cluster masses are consistent with the theoretical model but their distribution does not always cover the full simulated range as a result of observational selection. In particular the lower mass systems are underrepresented in the data. The lower panels of Figure \ref{fig:cl_obs} compare the $\sigcl$ evolution. We multiply the observed line-of-sight $\sigma_{cl}$ by $\sqrt{3}$ for comparison with the 3D simulated cluster velocity dispersions. The observed $\sigcl$s are consistent with the theoretical model because the scaling relation between cluster mass and velocity dispersion is similar in the data and in the simulations (Figure \ref{fig:msigma}). 

%========================================
%========================================

\begin{figure*}
\centering
\includegraphics[scale=0.28]{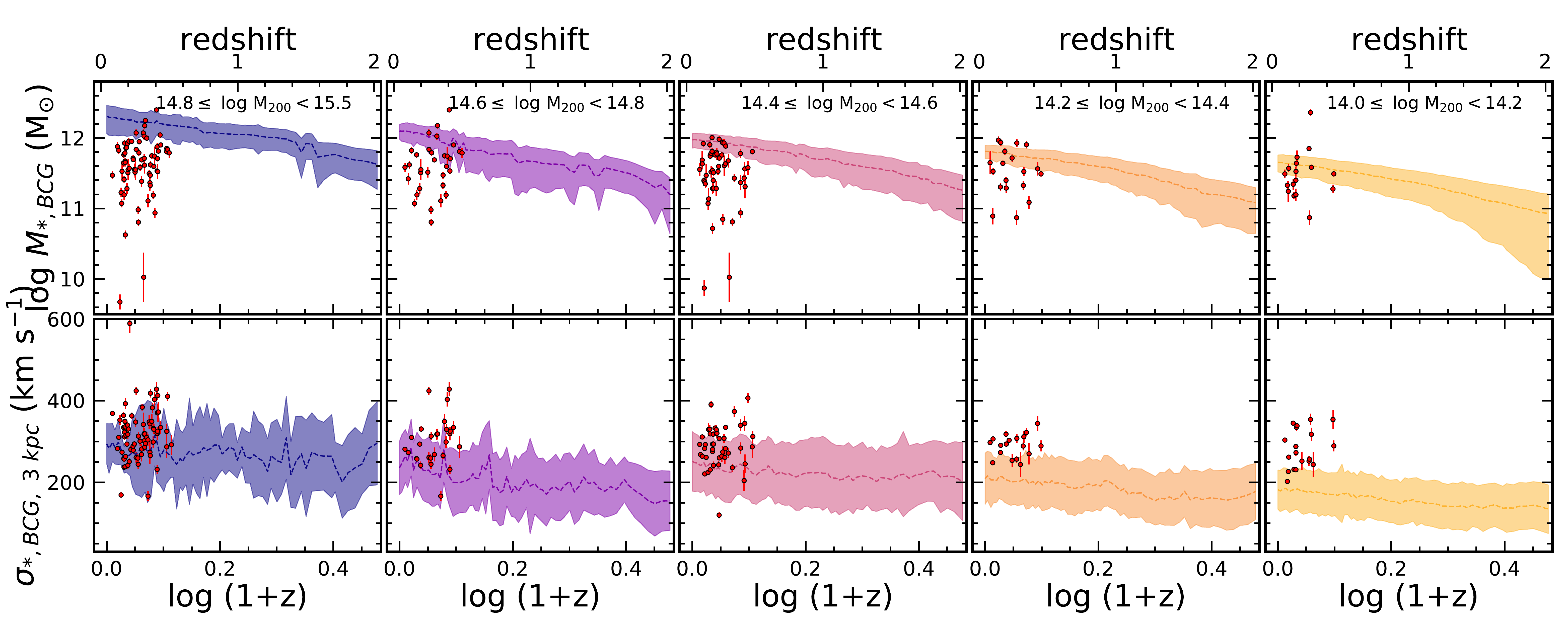}
\caption{The $M_{*}$ (upper panels) and the $\sigbcg$ (lower panels) of simulated BCGs (the dashed line and the shaded region) in five mass subsamples as a function of redshift. Red circles show the observed HeCS-omnibus clusters. }
\label{fig:bcg_obs}
\end{figure*}

We next compare the observed and simulated BCG properties. We use the BCGs of the observed cluster sample selected for Figure \ref{fig:cl_obs}. Figure \ref{fig:bcg_obs} displays $M_{*, BCG}$ and $\sigbcgc$ as a function of redshift. We also overlay the observed BCG properties from HeCS-omnibus (red circles, \citealp{Sohn20}). The HeCS-omnibus sample includes stellar mass estimates based on SDSS photometry and stellar velocity dispersions measured within a 3 kpc aperture (see details in \citealp{Sohn20}). For direct comparison with the observations, we use the simulated $\sigbcgc$ measured within a cylindrical volume that penetrates the center of BCGs within a 3 kpc aperture (see \citealp{Sohn22}). The substantial extension of the cylinder along the line-of-sight significantly reduces the impact of gravitational softening in this comparison between the observed and simulated velocity dispersions.

The $M_{*, BCG}$s of the observed BCGs overlap the models, but the observed $M_{*, BCG}$ of most clusters are smaller than the simulated $M_{*, BCG}$. The difference presumably results from challenges in estimating the BCG stellar mass. We use stellar mass estimates based on the SDSS composite Model (cModel) magnitude, which is unlikely to account completely for the extended BCG halo. Furthermore, the stellar mass from simulations is measured within the half-mass radius, not from the cModel magnitude. 

The distribution of $\sigbcg$ for massive clusters (log $M_{200} > 14.4$) is in better agreement with the simulations. However, the observed $\sigbcg$s in lower mass systems exceeds the simulated $\sigbcg$s. \citet{Sohn22} note this discrepancy based on comparison between the observed and simulated $\sigbcg \- \sigcl$ scaling relations. They suggest that the observed sample generally includes more massive systems with generally larger $\sigbcg$s as a result of observational selection. The difference in the $\sigbcg$ distributions needs to be further investigated with much larger observed samples of BCGs covering a wider redshift range and a larger cluster mass range at each redshift. Moreover, spatially resolved BCG velocity dispersion measurements are important for suppressing aperture effects.

%========================================
%========================================

Figure \ref{fig:cl_obs} and Figure \ref{fig:bcg_obs} promise interesting tests for future observations of high redshift ($z > 0.3$) clusters and BCGs. Current and future large spectroscopic surveys including DESI, 4MOST, MOONS, and Subaru/PFS will obtain dense spectroscopy for large numbers of high redshift cluster galaxies. Direct comparison between the cluster and BCG velocity dispersions derived from these surveys will be an immediate test of the simulations.

\citet{Sohn22} suggest a similar test based on comparison between observed and simulated dynamical scaling relations. \citet{Sohn22} selected the most massive cluster halos ($M_{200} > 10^{14}~\Msun$ at $z < 1$ snapshots rather than tracing the progenitors of massive cluster halos at $z = 0$. Tracing the progenitors requires dense observations at a succession of discrete redshifts. It is important that the observed samples are large enough and deep enough to select mass limited samples of possible progenitors at every redshift.

The impact of major mergers on the velocity dispersion evolution is another interesting testable prediction of the IllustrisTNG models. The BCG velocity dispersion increases abruptly when the BCG experiences a major merger. These precipitous changes in the velocity dispersion can be tested with IFU observations of hundreds of high redshift clusters selected to constitute  mass limited cluster samples covering a significant redshift range. Because the mergers are rare, the number of clusters must be large enough to catch the mergers in action and to outline any differences in merger frequency between high redshift and lower redshift samples. Deep photometric observations would provide additional confirmation of merging features associated with the BCGs. Taken together the spatially resolved spectroscopy and deep imaging would provide a picture of BCG evolution for direct comparison with the IllustrisTNG predictions.

%========================================
\section{Conclusion} \label{sec:conclusion}
%========================================

We investigate the evolution of massive clusters and their BCG at $z < 2$ based on the IllustrisTNG-300 cosmological simulation. We select 270 massive clusters with $M_{200} > 10^{14}~\Msun$ at $z = 0$. We trace the main progenitors of clusters and their BCGs in higher redshift snapshots using SubLink merger trees. 

We derive the physical properties of clusters and their BCGs from redshift snapshots for $z < 2$. We explore the sizes, masses and velocity dispersions of the systems: $R_{200}$, $M_{200}$ and $\sigcl$ for clusters and $R_{*, h}$, $M_{*}$, and $\sigbcg$ for BCGs. These properties of clusters and BCGs generally increase as the universe ages. The redshift evolution of these properties fluctuates because the clusters and BCGs interact dynamically with surrounding systems. 

We explore various scaling relations and their redshift evolution. The $M_{200} \- \sigcl$ relations have a consistent slope of $\sim 0.33$ at $z < 2$. The $M_{200} \- \sigbcg$ relations have a relatively shallower slopes. Interestingly, $\sigbcg$s are comparable with $\sigcl$ at high redshift ($z > 1$), while $\sigbcg$s are significantly smaller than $\sigcl$ at low redshift. The ratio $(\sigbcg / \sigcl)$ is correlated with $\sigcl$ over the redshift range we explore. The $(\sigbcg / \sigcl)$ decreases as a function of $\sigcl$. The relation is steeper at higher redshift. The underlying large $\sigbcg / \sigcl$ of high redshift cluster promises an interesting test of these models with future spectroscopic observations. 

We also trace the evolution of individual clusters and BCGs. We show that the $\sigbcg / \sigcl$ generally decreases as a function of $\sigcl$ (and redshift). In some cases, the $\sigbcg / \sigcl$ evolutions show abrupt changes because the BCG velocity dispersion is inflated as a result of major mergers between the BCGs and surrounding subhalos. The trends in $(\sigbcg / \sigcl) \- \sigcl$ and $(\sigbcg / \sigcl) \- z$ relations are insensitive to the cluster halo mass.

We explore the impact of aperture size on the dynamical scaling relations. Depending on the aperture size we use to measure $\sigbcg$s, the slopes of the dynamical scaling relations vary, but the general trends do not change. We note that $\sigbcg$s measured within a tight aperture of 3 kpc at high redshift ($z > 1$) is not a good physical proxy because of the complicated dynamical nature of BCGs at high redshift. A large aperture ($> 10$ kpc) measurement at higher redshift is required to understand the BCG dynamical properties. 

Comparison between the $M_{*} \- M_{200}$ relation and the $\sigbcg \- M_{200}$ relation demonstrates that $\sigbcg$ is a useful BCG mass proxy in tandem with $M_{*}$. Testing the scaling relations based on $\sigbcg$ provides an independent probe of the co-evolution of BCGs and their host clusters. Investigation on these relations also tests the prescription for quenching by supermassive black holes that is implemented in the IllustrisTNG simulation (e.g., \citealp{Vogelsberger13, Weinberger18}). The impact of this quenching only affects the mass and velocity dispersion growth of the BCGs. The mass and velocity dispersion evolutions of the host clusters are insensitive to  feedback. Thus the relative evolution of clusters and their BCGs provides an independent test of feedback models.

The redshift evolution of cluster and BCG properties we derive from simulations provide theoretical guidance for future observations. The models overlap the observed properties of clusters and BCGs from the HeCS-omnibus survey covering $z \lesssim 0.3$. Future much larger spectroscopic surveys, including IFU observations, will enable more extensive tests of the behavior higher redshift systems against the model predictions. These tests include examination of the impact of major mergers on BCG evolution as a function of cosmological epoch.

\acknowledgements
We thank our anonymous referee for helpful and insightful comments that improve this manuscript. We thank Antonaldo Diaferio, Ken Rines, and Ivana Damjanov for helpful discussions. This work was supported by the New Faculty Startup Fund from Seoul National University. J.S. is supported by the CfA Fellowship. M.J.G. acknowledges the Smithsonian Institution for support. MV acknowledges support through NASA ATP 19-ATP19-0019, 19-ATP19-0020, 19-ATP19-0167, and NSF grants AST-1814053, AST-1814259, AST-1909831, AST-2007355 and AST-2107724. MV also acknowledges support from a MIT RSC award, the Alfred P. Sloan Foundation, and by NASA ATP grant NNX17AG29G. I.D. acknowledges the support of the Canada Research Chair Program and the Natural Sciences and Engineering Research Council of Canada (NSERC, funding reference number RGPIN-2018-05425).

All of the primary TNG simulations have been run on the Cray XC40 Hazel Hen supercomputer at the High Performance Computing Center Stuttgart (HLRS) in Germany. They have been made possible by the Gauss Centre for Supercomputing (GCS) large-scale project proposals GCS-ILLU and GCS-DWAR. GCS is the alliance of the three national supercomputing centres HLRS (Universitaet Stuttgart), JSC (Forschungszentrum Julich), and LRZ (Bayerische Akademie der Wissenschaften), funded by the German Federal Ministry of Education and Research (BMBF) and the German State Ministries for Research of Baden-Wuerttemberg (MWK), Bayern (StMWFK) and Nordrhein-Westfalen (MIWF). Further simulations were run on the Hydra and Draco supercomputers at the Max Planck Computing and Data Facility (MPCDF, formerly known as RZG) in Garching near Munich, in addition to the Magny system at HITS in Heidelberg. Additional computations were carried out on the Odyssey2 system supported by the FAS Division of Science, Research Computing Group at Harvard University, and the Stampede supercomputer at the Texas Advanced Computing Center through the XSEDE project AST140063.
\clearpage

%========================================
% bibliography
%========================================
\bibliographystyle{aasjournal}
\bibliography{ms}

\begin{thebibliography}{}
\expandafter\ifx\csname natexlab\endcsname\relax\def\natexlab#1{#1}\fi
\providecommand{\url}[1]{\href{#1}{#1}}
\providecommand{\dodoi}[1]{doi:~\href{http://doi.org/#1}{\nolinkurl{#1}}}
\providecommand{\doeprint}[1]{\href{http://ascl.net/#1}{\nolinkurl{http://ascl.net/#1}}}
\providecommand{\doarXiv}[1]{\href{https://arxiv.org/abs/#1}{\nolinkurl{https://arxiv.org/abs/#1}}}

\bibitem[{{Beers} {et~al.}(1990){Beers}, {Flynn}, \& {Gebhardt}}]{Beers90}
{Beers}, T.~C., {Flynn}, K., \& {Gebhardt}, K. 1990, \aj, 100, 32,
  \dodoi{10.1086/115487}

\bibitem[{{Behroozi} {et~al.}(2019){Behroozi}, {Wechsler}, {Hearin}, \&
  {Conroy}}]{Behroozi19}
{Behroozi}, P., {Wechsler}, R.~H., {Hearin}, A.~P., \& {Conroy}, C. 2019,
  \mnras, 488, 3143, \dodoi{10.1093/mnras/stz1182}

\bibitem[{{Bernardi}(2009)}]{Bernardi09}
{Bernardi}, M. 2009, \mnras, 395, 1491,
  \dodoi{10.1111/j.1365-2966.2009.14601.x}

\bibitem[{{Bose} \& {Loeb}(2021)}]{Bose21}
{Bose}, S., \& {Loeb}, A. 2021, \apj, 912, 114,
  \dodoi{10.3847/1538-4357/abec77}

\bibitem[{{Campbell} {et~al.}(2017){Campbell}, {Frenk}, {Jenkins}, {Eke},
  {Navarro}, {Sawala}, {Schaller}, {Fattahi}, {Oman}, \& {Theuns}}]{Campbell17}
{Campbell}, D. J.~R., {Frenk}, C.~S., {Jenkins}, A., {et~al.} 2017, \mnras,
  469, 2335, \dodoi{10.1093/mnras/stx975}

\bibitem[{{De Boni} {et~al.}(2016){De Boni}, {Serra}, {Diaferio}, {Giocoli}, \&
  {Baldi}}]{DeBoni16}
{De Boni}, C., {Serra}, A.~L., {Diaferio}, A., {Giocoli}, C., \& {Baldi}, M.
  2016, \apj, 818, 188, \dodoi{10.3847/0004-637X/818/2/188}

\bibitem[{{DeMaio} {et~al.}(2020){DeMaio}, {Gonzalez}, {Zabludoff}, {Zaritsky},
  {Aldering}, {Brodwin}, {Connor}, {Donahue}, {Hayden}, {Mulchaey},
  {Perlmutter}, \& {Stanford}}]{DeMaio20}
{DeMaio}, T., {Gonzalez}, A.~H., {Zabludoff}, A., {et~al.} 2020, \mnras, 491,
  3751, \dodoi{10.1093/mnras/stz3236}

\bibitem[{{Diaferio}(1999)}]{Diaferio99}
{Diaferio}, A. 1999, \mnras, 309, 610, \dodoi{10.1046/j.1365-8711.1999.02864.x}

\bibitem[{{Diaferio} \& {Geller}(1997)}]{Diaferio97}
{Diaferio}, A., \& {Geller}, M.~J. 1997, \apj, 481, 633, \dodoi{10.1086/304075}

\bibitem[{{Dolag} {et~al.}(2010){Dolag}, {Murante}, \& {Borgani}}]{Dolag10}
{Dolag}, K., {Murante}, G., \& {Borgani}, S. 2010, \mnras, 405, 1544,
  \dodoi{10.1111/j.1365-2966.2010.16583.x}

\bibitem[{{Erfanianfar} {et~al.}(2019){Erfanianfar}, {Finoguenov}, {Furnell},
  {Popesso}, {Biviano}, {Wuyts}, {Collins}, {Mirkazemi}, {Comparat},
  {Khosroshahi}, {Nandra}, {Capasso}, {Rykoff}, {Wilman}, {Merloni}, {Clerc},
  {Salvato}, {Chitham}, {Kelvin}, {Gozaliasl}, {Weijmans}, {Brownstein},
  {Egami}, {Pereira}, {Schneider}, {Kirkpatrick}, {Damsted}, \&
  {Kukkola}}]{Erfanianfar19}
{Erfanianfar}, G., {Finoguenov}, A., {Furnell}, K., {et~al.} 2019, \aap, 631,
  A175, \dodoi{10.1051/0004-6361/201935375}

\bibitem[{{Fakhouri} {et~al.}(2010){Fakhouri}, {Ma}, \&
  {Boylan-Kolchin}}]{Fakhouri10}
{Fakhouri}, O., {Ma}, C.-P., \& {Boylan-Kolchin}, M. 2010, \mnras, 406, 2267,
  \dodoi{10.1111/j.1365-2966.2010.16859.x}

\bibitem[{{Girelli} {et~al.}(2020){Girelli}, {Pozzetti}, {Bolzonella},
  {Giocoli}, {Marulli}, \& {Baldi}}]{Girelli20}
{Girelli}, G., {Pozzetti}, L., {Bolzonella}, M., {et~al.} 2020, \aap, 634,
  A135, \dodoi{10.1051/0004-6361/201936329}

\bibitem[{{Jung} {et~al.}(2022){Jung}, {Rennehan}, {Saeedzadeh}, {Babul},
  {Tremmel}, {Quinn}, {Ilani Loubser}, {O'Sullivan}, \& {Yi}}]{Jung22}
{Jung}, S.~L., {Rennehan}, D., {Saeedzadeh}, V., {et~al.} 2022, arXiv e-prints,
  arXiv:2203.00016.
\newblock \doarXiv{2203.00016}

\bibitem[{{Kim} {et~al.}(2017){Kim}, {Ko}, {Hwang}, {Edge}, {Lee}, {Lee}, \&
  {Jeong}}]{Kim17}
{Kim}, J.-W., {Ko}, J., {Hwang}, H.~S., {et~al.} 2017, \apj, 836, 105,
  \dodoi{10.3847/1538-4357/aa5b8e}

\bibitem[{{Kravtsov} \& {Borgani}(2012)}]{Kravtsov12}
{Kravtsov}, A.~V., \& {Borgani}, S. 2012, \araa, 50, 353,
  \dodoi{10.1146/annurev-astro-081811-125502}

\bibitem[{{Kravtsov} {et~al.}(2018){Kravtsov}, {Vikhlinin}, \&
  {Meshcheryakov}}]{Kravtsov18}
{Kravtsov}, A.~V., {Vikhlinin}, A.~A., \& {Meshcheryakov}, A.~V. 2018,
  Astronomy Letters, 44, 8, \dodoi{10.1134/S1063773717120015}

\bibitem[{{Lauer} {et~al.}(2014){Lauer}, {Postman}, {Strauss}, {Graves}, \&
  {Chisari}}]{Lauer14}
{Lauer}, T.~R., {Postman}, M., {Strauss}, M.~A., {Graves}, G.~J., \& {Chisari},
  N.~E. 2014, \apj, 797, 82, \dodoi{10.1088/0004-637X/797/2/82}

\bibitem[{{Lin} \& {Mohr}(2004)}]{Lin04}
{Lin}, Y.-T., \& {Mohr}, J.~J. 2004, \apj, 617, 879, \dodoi{10.1086/425412}

\bibitem[{{Lin} {et~al.}(2017){Lin}, {Hsieh}, {Lin}, {Oguri}, {Chen}, {Tanaka},
  {Chiu}, {Huang}, {Kodama}, {Leauthaud}, {More}, {Nishizawa}, {Bundy}, {Lin},
  \& {Miyazaki}}]{Lin17}
{Lin}, Y.-T., {Hsieh}, B.-C., {Lin}, S.-C., {et~al.} 2017, \apj, 851, 139,
  \dodoi{10.3847/1538-4357/aa9bf5}

\bibitem[{{Loubser} {et~al.}(2018){Loubser}, {Hoekstra}, {Babul}, \&
  {O'Sullivan}}]{Loubser18}
{Loubser}, S.~I., {Hoekstra}, H., {Babul}, A., \& {O'Sullivan}, E. 2018,
  \mnras, 477, 335, \dodoi{10.1093/mnras/sty498}

\bibitem[{{Marinacci} {et~al.}(2018){Marinacci}, {Vogelsberger}, {Pakmor},
  {Torrey}, {Springel}, {Hernquist}, {Nelson}, {Weinberger}, {Pillepich},
  {Naiman}, \& {Genel}}]{Marinacci18}
{Marinacci}, F., {Vogelsberger}, M., {Pakmor}, R., {et~al.} 2018, \mnras, 480,
  5113, \dodoi{10.1093/mnras/sty2206}

\bibitem[{{Marini} {et~al.}(2021){Marini}, {Borgani}, {Saro}, {Granato},
  {Ragone-Figueroa}, {Sartoris}, {Dolag}, {Murante}, {Ragagnin}, \&
  {Wang}}]{Marini21}
{Marini}, I., {Borgani}, S., {Saro}, A., {et~al.} 2021, \mnras, 507, 5780,
  \dodoi{10.1093/mnras/stab2518}

\bibitem[{{McBride} {et~al.}(2009){McBride}, {Fakhouri}, \& {Ma}}]{McBride09}
{McBride}, J., {Fakhouri}, O., \& {Ma}, C.-P. 2009, \mnras, 398, 1858,
  \dodoi{10.1111/j.1365-2966.2009.15329.x}

\bibitem[{{Moster} {et~al.}(2010){Moster}, {Somerville}, {Maulbetsch}, {van den
  Bosch}, {Macci{\`o}}, {Naab}, \& {Oser}}]{Moster10}
{Moster}, B.~P., {Somerville}, R.~S., {Maulbetsch}, C., {et~al.} 2010, \apj,
  710, 903, \dodoi{10.1088/0004-637X/710/2/903}

\bibitem[{{Naiman} {et~al.}(2018){Naiman}, {Pillepich}, {Springel},
  {Ramirez-Ruiz}, {Torrey}, {Vogelsberger}, {Pakmor}, {Nelson}, {Marinacci},
  {Hernquist}, {Weinberger}, \& {Genel}}]{Naiman18}
{Naiman}, J.~P., {Pillepich}, A., {Springel}, V., {et~al.} 2018, \mnras, 477,
  1206, \dodoi{10.1093/mnras/sty618}

\bibitem[{{Nelson} {et~al.}(2018){Nelson}, {Pillepich}, {Springel},
  {Weinberger}, {Hernquist}, {Pakmor}, {Genel}, {Torrey}, {Vogelsberger},
  {Kauffmann}, {Marinacci}, \& {Naiman}}]{Nelson18}
{Nelson}, D., {Pillepich}, A., {Springel}, V., {et~al.} 2018, \mnras, 475, 624,
  \dodoi{10.1093/mnras/stx3040}

\bibitem[{{Nelson} {et~al.}(2019){Nelson}, {Springel}, {Pillepich},
  {Rodriguez-Gomez}, {Torrey}, {Genel}, {Vogelsberger}, {Pakmor}, {Marinacci},
  {Weinberger}, {Kelley}, {Lovell}, {Diemer}, \& {Hernquist}}]{Nelson19}
{Nelson}, D., {Springel}, V., {Pillepich}, A., {et~al.} 2019, Computational
  Astrophysics and Cosmology, 6, 2, \dodoi{10.1186/s40668-019-0028-x}

\bibitem[{{Oliva-Altamirano} {et~al.}(2014){Oliva-Altamirano}, {Brough},
  {Lidman}, {Couch}, {Hopkins}, {Colless}, {Taylor}, {Robotham},
  {Gunawardhana}, {Ponman}, {Baldry}, {Bauer}, {Bland-Hawthorn}, {Cluver},
  {Cameron}, {Conselice}, {Driver}, {Edge}, {Graham}, {van Kampen},
  {Lara-L{\'o}pez}, {Liske}, {L{\'o}pez-S{\'a}nchez}, {Loveday}, {Mahajan},
  {Peacock}, {Phillipps}, {Pimbblet}, \& {Sharp}}]{OlivaAltamirano14}
{Oliva-Altamirano}, P., {Brough}, S., {Lidman}, C., {et~al.} 2014, \mnras, 440,
  762, \dodoi{10.1093/mnras/stu277}

\bibitem[{{Pillepich} {et~al.}(2018){Pillepich}, {Nelson}, {Hernquist},
  {Springel}, {Pakmor}, {Torrey}, {Weinberger}, {Genel}, {Naiman}, {Marinacci},
  \& {Vogelsberger}}]{Pillepich18b}
{Pillepich}, A., {Nelson}, D., {Hernquist}, L., {et~al.} 2018, \mnras, 475,
  648, \dodoi{10.1093/mnras/stx3112}

\bibitem[{{Pizzardo} {et~al.}(2022){Pizzardo}, {Sohn}, {Geller}, {Diaferio}, \&
  {Rines}}]{Pizzardo22}
{Pizzardo}, M., {Sohn}, J., {Geller}, M.~J., {Diaferio}, A., \& {Rines}, K.
  2022, \apj, 927, 26, \dodoi{10.3847/1538-4357/ac5029}

\bibitem[{{Pizzardo} {et~al.}(2021){Pizzardo}, {Di Gioia}, {Diaferio}, {De
  Boni}, {Serra}, {Geller}, {Sohn}, {Rines}, \& {Baldi}}]{Pizzardo21}
{Pizzardo}, M., {Di Gioia}, S., {Diaferio}, A., {et~al.} 2021, \aap, 646, A105,
  \dodoi{10.1051/0004-6361/202038481}

\bibitem[{{Planck Collaboration} {et~al.}(2016){Planck Collaboration}, {Ade},
  {Aghanim}, {Arnaud}, {Ashdown}, {Aumont}, {Baccigalupi}, {Banday},
  {Barreiro}, {Bartlett}, {Bartolo}, {Battaner}, {Battye}, {Benabed},
  {Beno{\^\i}t}, {Benoit-L{\'e}vy}, {Bernard}, {Bersanelli}, {Bielewicz},
  {Bock}, {Bonaldi}, {Bonavera}, {Bond}, {Borrill}, {Bouchet}, {Boulanger},
  {Bucher}, {Burigana}, {Butler}, {Calabrese}, {Cardoso}, {Catalano},
  {Challinor}, {Chamballu}, {Chary}, {Chiang}, {Chluba}, {Christensen},
  {Church}, {Clements}, {Colombi}, {Colombo}, {Combet}, {Coulais}, {Crill},
  {Curto}, {Cuttaia}, {Danese}, {Davies}, {Davis}, {de Bernardis}, {de Rosa},
  {de Zotti}, {Delabrouille}, {D{\'e}sert}, {Di Valentino}, {Dickinson},
  {Diego}, {Dolag}, {Dole}, {Donzelli}, {Dor{\'e}}, {Douspis}, {Ducout},
  {Dunkley}, {Dupac}, {Efstathiou}, {Elsner}, {En{\ss}lin}, {Eriksen},
  {Farhang}, {Fergusson}, {Finelli}, {Forni}, {Frailis}, {Fraisse},
  {Franceschi}, {Frejsel}, {Galeotta}, {Galli}, {Ganga}, {Gauthier}, {Gerbino},
  {Ghosh}, {Giard}, {Giraud-H{\'e}raud}, {Giusarma}, {Gjerl{\o}w},
  {Gonz{\'a}lez-Nuevo}, {G{\'o}rski}, {Gratton}, {Gregorio}, {Gruppuso},
  {Gudmundsson}, {Hamann}, {Hansen}, {Hanson}, {Harrison}, {Helou},
  {Henrot-Versill{\'e}}, {Hern{\'a}ndez-Monteagudo}, {Herranz}, {Hildebrandt},
  {Hivon}, {Hobson}, {Holmes}, {Hornstrup}, {Hovest}, {Huang}, {Huffenberger},
  {Hurier}, {Jaffe}, {Jaffe}, {Jones}, {Juvela}, {Keih{\"a}nen}, {Keskitalo},
  {Kisner}, {Kneissl}, {Knoche}, {Knox}, {Kunz}, {Kurki-Suonio}, {Lagache},
  {L{\"a}hteenm{\"a}ki}, {Lamarre}, {Lasenby}, {Lattanzi}, {Lawrence}, {Leahy},
  {Leonardi}, {Lesgourgues}, {Levrier}, {Lewis}, {Liguori}, {Lilje},
  {Linden-V{\o}rnle}, {L{\'o}pez-Caniego}, {Lubin}, {Mac{\'\i}as-P{\'e}rez},
  {Maggio}, {Maino}, {Mandolesi}, {Mangilli}, {Marchini}, {Maris}, {Martin},
  {Martinelli}, {Mart{\'\i}nez-Gonz{\'a}lez}, {Masi}, {Matarrese}, {McGehee},
  {Meinhold}, {Melchiorri}, {Melin}, {Mendes}, {Mennella}, {Migliaccio},
  {Millea}, {Mitra}, {Miville-Desch{\^e}nes}, {Moneti}, {Montier}, {Morgante},
  {Mortlock}, {Moss}, {Munshi}, {Murphy}, {Naselsky}, {Nati}, {Natoli},
  {Netterfield}, {N{\o}rgaard-Nielsen}, {Noviello}, {Novikov}, {Novikov},
  {Oxborrow}, {Paci}, {Pagano}, {Pajot}, {Paladini}, {Paoletti}, {Partridge},
  {Pasian}, {Patanchon}, {Pearson}, {Perdereau}, {Perotto}, {Perrotta},
  {Pettorino}, {Piacentini}, {Piat}, {Pierpaoli}, {Pietrobon}, {Plaszczynski},
  {Pointecouteau}, {Polenta}, {Popa}, {Pratt}, {Pr{\'e}zeau}, {Prunet},
  {Puget}, {Rachen}, {Reach}, {Rebolo}, {Reinecke}, {Remazeilles}, {Renault},
  {Renzi}, {Ristorcelli}, {Rocha}, {Rosset}, {Rossetti}, {Roudier},
  {Rouill{\'e} d'Orfeuil}, {Rowan-Robinson}, {Rubi{\~n}o-Mart{\'\i}n},
  {Rusholme}, {Said}, {Salvatelli}, {Salvati}, {Sandri}, {Santos},
  {Savelainen}, {Savini}, {Scott}, {Seiffert}, {Serra}, {Shellard}, {Spencer},
  {Spinelli}, {Stolyarov}, {Stompor}, {Sudiwala}, {Sunyaev}, {Sutton},
  {Suur-Uski}, {Sygnet}, {Tauber}, {Terenzi}, {Toffolatti}, {Tomasi},
  {Tristram}, {Trombetti}, {Tucci}, {Tuovinen}, {T{\"u}rler}, {Umana},
  {Valenziano}, {Valiviita}, {Van Tent}, {Vielva}, {Villa}, {Wade}, {Wandelt},
  {Wehus}, {White}, {White}, {Wilkinson}, {Yvon}, {Zacchei}, \&
  {Zonca}}]{Planck16}
{Planck Collaboration}, {Ade}, P.~A.~R., {Aghanim}, N., {et~al.} 2016, \aap,
  594, A13, \dodoi{10.1051/0004-6361/201525830}

\bibitem[{{Ragone-Figueroa} {et~al.}(2018){Ragone-Figueroa}, {Granato},
  {Ferraro}, {Murante}, {Biffi}, {Borgani}, {Planelles}, \&
  {Rasia}}]{RagoneFigueroa18}
{Ragone-Figueroa}, C., {Granato}, G.~L., {Ferraro}, M.~E., {et~al.} 2018,
  \mnras, 479, 1125, \dodoi{10.1093/mnras/sty1639}

\bibitem[{{Remus} {et~al.}(2017){Remus}, {Dolag}, \& {Hoffmann}}]{Remus17}
{Remus}, R.-S., {Dolag}, K., \& {Hoffmann}, T. 2017, Galaxies, 5, 49,
  \dodoi{10.3390/galaxies5030049}

\bibitem[{{Rines} \& {Diaferio}(2006)}]{Rines06}
{Rines}, K., \& {Diaferio}, A. 2006, \aj, 132, 1275, \dodoi{10.1086/506017}

\bibitem[{{Rines} {et~al.}(2013){Rines}, {Geller}, {Diaferio}, \&
  {Kurtz}}]{Rines13}
{Rines}, K., {Geller}, M.~J., {Diaferio}, A., \& {Kurtz}, M.~J. 2013, \apj,
  767, 15, \dodoi{10.1088/0004-637X/767/1/15}

\bibitem[{{Rines} {et~al.}(2016){Rines}, {Geller}, {Diaferio}, \&
  {Hwang}}]{Rines16}
{Rines}, K.~J., {Geller}, M.~J., {Diaferio}, A., \& {Hwang}, H.~S. 2016, \apj,
  819, 63, \dodoi{10.3847/0004-637X/819/1/63}

\bibitem[{{Rines} {et~al.}(2018){Rines}, {Geller}, {Diaferio}, {Hwang}, \&
  {Sohn}}]{Rines18}
{Rines}, K.~J., {Geller}, M.~J., {Diaferio}, A., {Hwang}, H.~S., \& {Sohn}, J.
  2018, \apj, 862, 172, \dodoi{10.3847/1538-4357/aacd49}

\bibitem[{{Rodriguez-Gomez} {et~al.}(2015){Rodriguez-Gomez}, {Genel},
  {Vogelsberger}, {Sijacki}, {Pillepich}, {Sales}, {Torrey}, {Snyder},
  {Nelson}, {Springel}, {Ma}, \& {Hernquist}}]{RodriguezGomez15}
{Rodriguez-Gomez}, V., {Genel}, S., {Vogelsberger}, M., {et~al.} 2015, \mnras,
  449, 49, \dodoi{10.1093/mnras/stv264}

\bibitem[{{Serra} \& {Diaferio}(2013)}]{Serra13}
{Serra}, A.~L., \& {Diaferio}, A. 2013, \apj, 768, 116,
  \dodoi{10.1088/0004-637X/768/2/116}

\bibitem[{{Sohn} {et~al.}(2020){Sohn}, {Geller}, {Diaferio}, \&
  {Rines}}]{Sohn20}
{Sohn}, J., {Geller}, M.~J., {Diaferio}, A., \& {Rines}, K.~J. 2020, \apj, 891,
  129, \dodoi{10.3847/1538-4357/ab6e6a}

\bibitem[{{Sohn} {et~al.}(2021){Sohn}, {Geller}, {Hwang}, {Diaferio}, {Rines},
  \& {Utsumi}}]{Sohn21}
{Sohn}, J., {Geller}, M.~J., {Hwang}, H.~S., {et~al.} 2021, \apj, 923, 143,
  \dodoi{10.3847/1538-4357/ac29c3}

\bibitem[{{Sohn} {et~al.}(2022){Sohn}, {Geller}, {Vogelsberger}, \&
  {Danjanov}}]{Sohn22}
{Sohn}, J., {Geller}, M.~J., {Vogelsberger}, M., \& {Danjanov}, I. 2022, arXiv
  e-prints, arXiv:2201.08853.
\newblock \doarXiv{2201.08853}

\bibitem[{{Sohn} {et~al.}(2017){Sohn}, {Geller}, {Zahid}, {Fabricant},
  {Diaferio}, \& {Rines}}]{Sohn17a}
{Sohn}, J., {Geller}, M.~J., {Zahid}, H.~J., {et~al.} 2017, \apjs, 229, 20,
  \dodoi{10.3847/1538-4365/aa653e}

\bibitem[{{Springel} {et~al.}(2018){Springel}, {Pakmor}, {Pillepich},
  {Weinberger}, {Nelson}, {Hernquist}, {Vogelsberger}, {Genel}, {Torrey},
  {Marinacci}, \& {Naiman}}]{Springel18}
{Springel}, V., {Pakmor}, R., {Pillepich}, A., {et~al.} 2018, \mnras, 475, 676,
  \dodoi{10.1093/mnras/stx3304}

\bibitem[{{van den Bosch}(2002)}]{vandenBosch02}
{van den Bosch}, F.~C. 2002, \mnras, 331, 98,
  \dodoi{10.1046/j.1365-8711.2002.05171.x}

\bibitem[{{Vogelsberger} {et~al.}(2013){Vogelsberger}, {Genel}, {Sijacki},
  {Torrey}, {Springel}, \& {Hernquist}}]{Vogelsberger13}
{Vogelsberger}, M., {Genel}, S., {Sijacki}, D., {et~al.} 2013, \mnras, 436,
  3031, \dodoi{10.1093/mnras/stt1789}

\bibitem[{{Vogelsberger} {et~al.}(2020){Vogelsberger}, {Marinacci}, {Torrey},
  \& {Puchwein}}]{Vogelsberger20}
{Vogelsberger}, M., {Marinacci}, F., {Torrey}, P., \& {Puchwein}, E. 2020,
  Nature Reviews Physics, 2, 42, \dodoi{10.1038/s42254-019-0127-2}

\bibitem[{{Vogelsberger} {et~al.}(2014{\natexlab{a}}){Vogelsberger}, {Genel},
  {Springel}, {Torrey}, {Sijacki}, {Xu}, {Snyder}, {Bird}, {Nelson}, \&
  {Hernquist}}]{Vogelsberger14a}
{Vogelsberger}, M., {Genel}, S., {Springel}, V., {et~al.} 2014{\natexlab{a}},
  \nat, 509, 177, \dodoi{10.1038/nature13316}

\bibitem[{{Vogelsberger} {et~al.}(2014{\natexlab{b}}){Vogelsberger}, {Genel},
  {Springel}, {Torrey}, {Sijacki}, {Xu}, {Snyder}, {Nelson}, \&
  {Hernquist}}]{Vogelsberger14b}
---. 2014{\natexlab{b}}, \mnras, 444, 1518, \dodoi{10.1093/mnras/stu1536}

\bibitem[{{Von Der Linden} {et~al.}(2007){Von Der Linden}, {Best}, {Kauffmann},
  \& {White}}]{vonderLinden07}
{Von Der Linden}, A., {Best}, P.~N., {Kauffmann}, G., \& {White}, S. D.~M.
  2007, \mnras, 379, 867, \dodoi{10.1111/j.1365-2966.2007.11940.x}

\bibitem[{{Weinberger} {et~al.}(2018){Weinberger}, {Springel}, {Pakmor},
  {Nelson}, {Genel}, {Pillepich}, {Vogelsberger}, {Marinacci}, {Naiman},
  {Torrey}, \& {Hernquist}}]{Weinberger18}
{Weinberger}, R., {Springel}, V., {Pakmor}, R., {et~al.} 2018, \mnras, 479,
  4056, \dodoi{10.1093/mnras/sty1733}

\bibitem[{{Wen} \& {Han}(2018)}]{Wen18}
{Wen}, Z.~L., \& {Han}, J.~L. 2018, \mnras, 481, 4158,
  \dodoi{10.1093/mnras/sty2533}

\bibitem[{{Zhao} {et~al.}(2009){Zhao}, {Jing}, {Mo}, \& {B{\"o}rner}}]{Zhao09}
{Zhao}, D.~H., {Jing}, Y.~P., {Mo}, H.~J., \& {B{\"o}rner}, G. 2009, \apj, 707,
  354, \dodoi{10.1088/0004-637X/707/1/354}

\end{thebibliography}

\end{document}